\documentclass[fleqn,usenatbib]{mnras}

\usepackage{newtxtext,newtxmath}

\usepackage[T1]{fontenc}
\usepackage{ae,aecompl}

\usepackage{graphicx}	
\usepackage{amsmath}	
\usepackage{amssymb}	
\usepackage{gensymb}

\title{SMASHing the low surface brightness SMC}

\author[Pol Massana et. al.]{Pol Massana,$^{1}$\thanks{e-mail: p.massana@surrey.ac.uk}
Noelia E. D. No\"{e}l,$^{1}$
David L. Nidever,$^{2}$
Denis Erkal,$^{1}$
\newauthor
Thomas J. L. de Boer,$^{3}$
Yumi Choi,$^{4,\, 2}$
Steven R. Majewski,$^{5}$
Knut Olsen,$^{6}$
\newauthor
Antonela Monachesi,$^{7, \, 8}$
Carme Gallart,$^{9, \, 10}$
Roeland P. van der Marel,$^{4, \, 11}$
\newauthor
Tom\'{a}s Ruiz-Lara,$^{9, \, 10}$
Dennis Zaritsky,$^{12}$
Nicolas F. Martin,$^{13, \, 14}$
Ricardo R. Mu\~noz,$^{15}$
\newauthor
Maria-Rosa L. Cioni,$^{16}$
Cameron P. M. Bell,$^{16}$
Eric F. Bell,$^{17}$
Guy S. Stringfellow,$^{18}$
\newauthor
Vasily Belokurov,$^{19}$
Matteo Monelli,$^{9, \, 10}$
Alistair R. Walker,$^{20}$
\newauthor
David Mart\'{i}nez-Delgado,$^{21}$
A. Katherina Vivas,$^{20}$
Blair C. Conn$^{22}$
\\
Affiliations are listed at the end of the paper
}

\date{Accepted XXX. Received YYY; in original form ZZZ}

\pubyear{2020}

\begin{document}
\label{firstpage}
\pagerange{\pageref{firstpage}--\pageref{lastpage}}
\maketitle

\newpage

\begin{abstract}
The periphery of the Small Magellanic Cloud (SMC) can unlock important information regarding galaxy formation and evolution in interacting systems. Here, we present a detailed study of the extended stellar structure of the SMC using deep colour-magnitude diagrams (CMDs), obtained as part of the Survey of the MAgellanic Stellar History (SMASH). Special care was taken in the decontamination of our data from MW foreground stars, including from foreground globular clusters NGC 362 and 47 Tuc. We derived the SMC surface brightness using a ``conservative'' approach from which we calculated the general parameters of the SMC, finding a staggered surface brightness profile. We also traced the fainter outskirts by constructing a stellar density profile. This approach, based on stellar counts of the oldest main sequence turn-off (MSTO) stars, uncovered a tidally disrupted stellar feature that reaches as far out as 12 degrees from the SMC centre. We also serendipitously found a faint feature of unknown origin located at $\sim 14$ degrees from the centre of the SMC and that we tentatively associated to a more distant structure. We compared our results to in-house simulations of a $1\times10^{9} M_\odot$ SMC, finding that its elliptical shape can be explained by its tidal disruption under the combined presence of the MW and the LMC. Finally, we found that the older stellar populations show a smooth profile while the younger component presents a jump in the density followed by a flat profile, confirming the heavily disturbed nature of the SMC.
\end{abstract}




\section{Introduction}
One of the most intriguing puzzles in modern astrophysics is to understand how galaxies form and evolve. There are two main approaches to study the processes involved in the formation and evolution of galaxies. The first employs lookback time studies to observe the progenitors of present day galaxies at progressively higher redshifts (e.g. the Cosmic Assembly Near-infrared Deep Extragalactic Survey [CANDELS], \citealt{Grogin2011}). The opportunity to study a large number of galaxies, of different morphological types and environments, has the advantage of avoiding the stochasticity inherent to small samples. However, this approach is indirect because in order to understand a present day galaxy, we must assume we can identify appropriate progenitors at different redshifts, a procedure fraught with potential biases. 
The second technique consists in revealing the formation and evolution of nearby galaxies through detailed chemical and dynamical studies of individual stars. This approach, known as ``Galactic Archaeology'', has the drawback that there are only a few dozen galaxies available to study in detail in the nearby Universe. In spite of the latter, there is now data of unprecedented quality from ground-based (e.g. SMASH, \citealt{Nidever2017}; VMC, \citealt{Cioni2011}; PAndAS, \citealt{McConnachie2009}; DELVE, \citealt{Mau2020}) and space-based (LCID collaboration, \citealt{Gallart2015}; \textit{Gaia}, \citealt{Gaia2018A}) telescopes that make the field of Galactic Archaeology increasingly exciting.

Our Local Group constitutes an excellent workplace to perform Galactic Archaeology because many clues about galaxy assembly and evolution processes persist --- often as coherent substructures --- in the faint outskirts of galaxies (\citealt{Bekki2005}; \citealt{Read2006c}; \citealt{Munoz2006a}; \citealt{Harmsen2017}; \citealt{Kallivayalil2018}; \citealt{Mackey2018}; \citealt{Belokurov2019}).
Hence, studying the dynamically unrelaxed
edges of nearby galaxies is particularly effective.
Although the extreme extents of galaxies are very faint (e.g. $\mu \approx 27$ mag/arcsec$^{2}$, \citealt{Gallart2004}; $\mu \approx 34$ mag /arcsec$^2$, \citealt{Nidever2019}), the Local Group galaxies offer the opportunity to study these extremely low surface brightness peripheries through direct observation of individually resolved stars.

Due to their interesting nature and close proximity, our closest interacting neighbours, the Small and Large Magellanic Clouds (SMC/LMC), are arguably the best laboratories at hand to study faint outskirts using resolved stellar populations. 
The Magellanic Clouds (MCs) are the largest and most massive dwarf galaxies under strong gravitational influence of the Milky Way (MW), with diameters of $\sim18$ kpc for the LMC \citep{Nidever2019} and $\sim11$ kpc for the SMC \citep{Nidever2011}, and halo masses of $\sim 10^{11}$M$_{\odot}$ for the LMC \citep{Kallivayalil2013,Penarrubia:2016,Erkal2019b} and $\sim 10^{9}$M$_{\odot}$ for the SMC \citep{DiTeodoro2019}. The MCs show compelling evidence of mutual interaction as unravelled by: (i) the kinematically distinct population found in the LMC that originated in the SMC \citep{Olsen2011}, (ii) the  stellar and gaseous Magellanic Bridge connecting both galaxies (\citealt{Hindman1963}, \citealt{Nidever2013}, \citealt{Noel2013a}, \citealt{Noel2015}, \citealt{Carrera2017}), (iii) the RR Lyrae overdensity located at $\sim5$ kpc below the main ridgeline of the Magellanic Bridge  \citep{Belokurov2017}, and (iv) the 200 degree, purely gaseous Magellanic Stream and Leading Arm features (\citealt{Mathewson1974}, \citealt{Nidever2010}, \citealt{Putman1998}).

Based on recent Hubble Space Telescope (HST) observations, the MCs show a high velocity relative to the MW \citep{Kallivayalil2006a, Kallivayalil2013}, suggesting that they are most likely on their first infall to the MW \citep{Besla2007B}. The pair have actually been interacting with each other for several gigayears (Gyr), including a likely recent collision between the two MCs \citep{Besla2012}, thought to have produced the low surface brightness, asymmetric features in the LMC \citep{Besla2016}. \citet{Zivick2018} showed that this close encounter happened $147 \pm 33$ Myr ago with an impact parameter of $7.5 \pm 2.5$ kpc, a distance that falls within the stellar extent of each galaxy and well within both of their virial radii.

The SMC, located $\sim 63.4 \, \rm{kpc}$ from us \citep{Ripepi2016}, is more metal poor than its larger companion, the LMC (\citealt{Carrera2008a}, \citealt{Carrera2008b}), and has an older field stellar population \citep{Noel2007A} than its oldest globular cluster NGC 121 \citep{Glatt2008}, as well as a unique star formation history (SFH) (\citealt{Noel2009}). In addition, the SMC exhibits a very complex geometry, with indications of significant depths along several lines of sight (\citealt{Gardiner1991}; \citealt{DeGrijs2015}; \citealt{Scowcroft2016}; \citealt{Ripepi2017}; \citealt{Muraveva2018}), and evidence of two different structures along its eastern wing (\citealt{Nidever2013}; \citealt{Piatti2015}; \citealt{Subramanian2017}). These properties make the SMC a fascinating object for detailed study. 

\citet{Noel2007B} were the first to suggest that the SMC was larger than previously thought, with its intermediate age population extending as far as 6.5 kpc from the SMC centre.  \citet{Nidever2011} later found evidence for SMC stars present at a radius of $\sim$7.5 kpc.  More recently, the Dark Energy Survey \citep[DES;][]{Flaugher2005}, thanks to its depth and large coverage, allowed for the discovery of an overdensity in the outskirts of the SMC at a projected radius of 8 kpc \citep{Pieres2017}, for a distance modulus of $18.96$ mag. 

In the present work, we aim to shed light on the origin of the SMC's irregular shape, its initial mass, the distribution of its stellar content at different ages and to prove the existence of a faint population in its outskirts out to $\sim$8 kpc or beyond. We build from previous works, to carry out a study based on more comprehensive data using the Dark Energy Camera (DECam, \citealt{Flaugher2015}).  We present an analysis of the SMC based on colour-magnitude diagrams (CMDs) with an unprecedented combination of depth and areal coverage, obtained as part of the Survey of the MAgellanic Stellar History (SMASH; \citealt{Nidever2017}). These deep CMDs are used to obtain surface brightness profiles of the SMC and to construct a map of stellar counts across the SMC's main body.

This paper is organised as follows. In Section \ref{sec:data} we describe the main features of the SMASH survey and give an overview of the data reduction and preparation. In Section \ref{sec:CMDdecontamination} we discuss the process carried out to select SMC stars, we describe the procedure followed to  decontaminate stars from the MW halo as well as two MW globular clusters (GCs) present in the SMC foreground (47 Tuc and NGC 362), and we assess sources of uncertainty affecting our analysis. In Section \ref{sec:SB profile} we present the derived surface brightness profile of the SMC and discuss the implications. In Section \ref{sec: MSTO profile} we present a stellar density profile of the SMC outskirts. In Section \ref{sec:simulations} we compare our results to numerical simulations with the goal of understanding the origin of the SMC's irregular shape and initial mass. In Section \ref{sec: young vs. old} we analyse the distribution of stellar populations of different ages. Finally, in Section \ref{sec:conclusions} we present our conclusions.

\section{Data} \label{sec:data}

\subsection{SMASH Overview} \label{subsec:smash_overview}

SMASH is a photometric survey performed using the large field of view ($\sim$3 deg$^2$) of DECam \citep{Flaugher2015} installed on the Blanco 4-m telescope at CTIO (Cerro Tololo Inter-American Observatory). SMASH surveys a net area of $\sim 480 \, \mathrm{deg}^2$ distributed across $\sim 2400 \, \mathrm{deg}^2$ surrounding the MCs --- i.e., with a 20\% filling factor. SMASH data span the \textit{ugriz} filter system, with all fields reaching a depth of at least $g\sim 24$ mag, with some reaching as faint as $g\sim 26$ mag. 
The combined depth and areal coverage are the best to date for the MCs. 
SMASH was designed with the main goals of recovering the SFHs of the MCs and detecting faint stellar structures in their outskirts. These tasks require reaching well below the oldest main sequence  turn-off (MSTO), that, for the Clouds, is located at $g \approx 22$ mag. Figure \ref{fig:study_map} shows the SMASH coverage around the main bodies of the MCs depicted as green and black hexagons, mostly adjacent to the DES footprint shown as a shaded pink area. For more technical information about SMASH, including a complete description of the data reduction process as well as its Data Release 1, we refer the reader to \citet{Nidever2017}.

SMASH has already shed light on several aspects of the MCs. During the observation phase, \citet{Martin2015} identified a new satellite of the MW, Hydra II. \citet{Nidever2019} explored the outskirts of the LMC using  SMASH data down to a surface brightness of  $\mu \approx 34$ mag /arcsec$^2$. Using red clump (RC) stars, \citet{Choi2018a} constructed an accurate, large-scale, reddening map for the main body of the LMC, and discovered a new stellar warp of its outer disc. Using the same selection of stars, \citet{Choi2018b} detected a ring-like overdensity in the LMC disc. Meanwhile, \citet{Martin2016a} discovered SMASH-1, a faint GC disrupting in the outskirts of the LMC. \citet{Martinez-Delgado2019} used SMASH data to study a young shell of stars that, given its derived age, could have originated from the recent LMC-SMC encounter.

\subsection{Data Processing} \label{subsec:data_preparation}

\begin{figure}
\centering
\includegraphics[width=0.95\columnwidth]{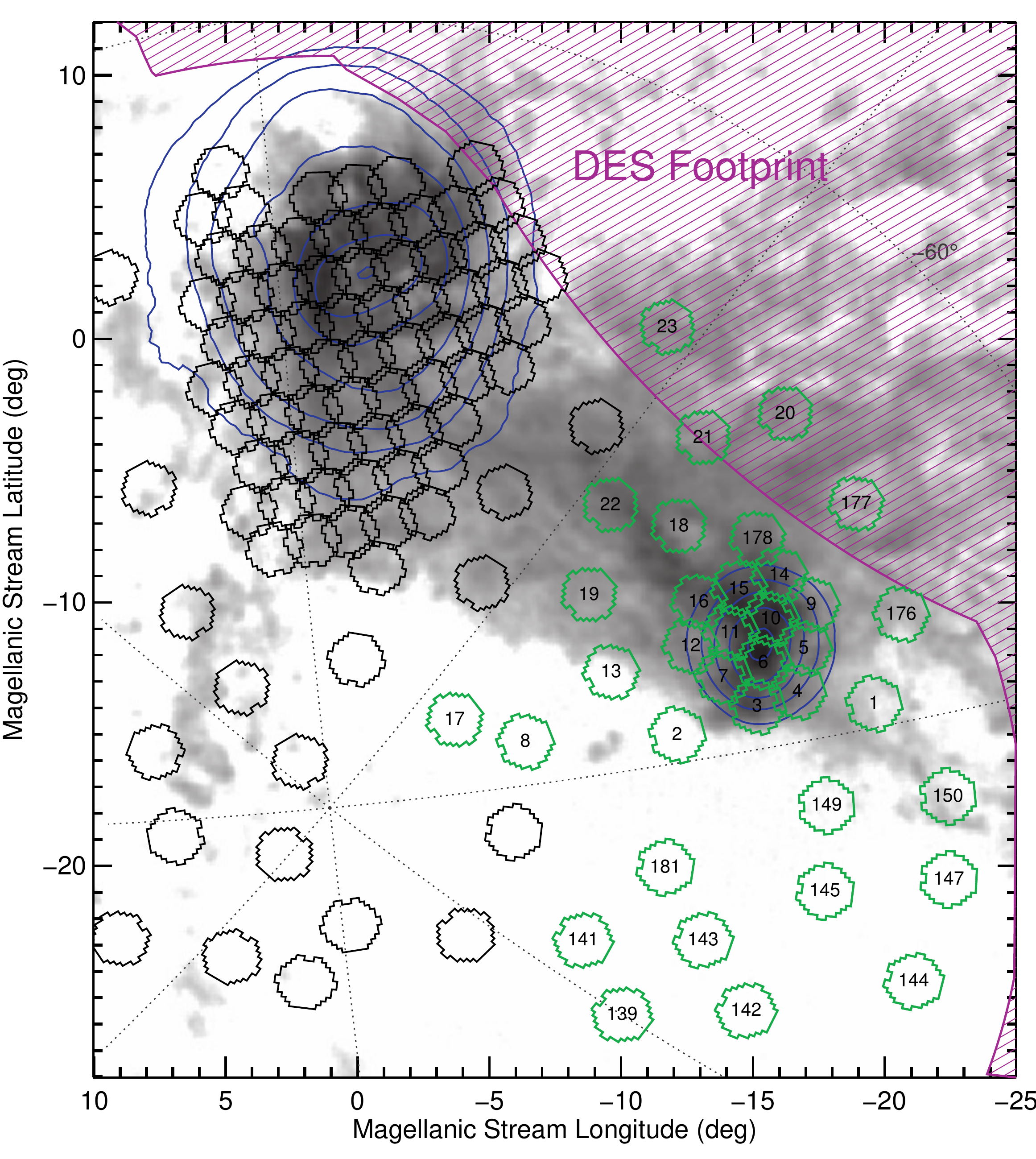}
\caption{Spatial distribution of the SMASH fields (shown as hexagonal polygons) across the MCs. The observed HI column density of the Magellanic Stream system is shown in grey scale \citep{Nidever2010}, while the dark blue contours  around the SMC and LMC represent RGB star counts from 2MASS \citep{Skrutskie2006}.
The green, numbered hexagons represent the 36 SMASH fields that are the focus of this study and that extend up to 15 degrees from the centre of the SMC. The DES footprint is represented by the purple shaded region.}
\label{fig:study_map}
\end{figure}

For the present study, we analysed 36 SMASH fields that cover a total area of $\sim 100 \, \mathrm{deg}^2$ in the sky, reaching as far as 15 degrees away from the SMC centre. These fields  are represented as green hexagons in Figure \ref{fig:study_map}. The field coverage in the central parts of the SMC is denser than in the outskirts,  allowing a complete mapping of the SFH of the SMC's inner regions (Massana et al. in prep). Figure \ref{fig:study_map}  also shows the LMC coverage, which extends well into its outskirts, represented as black hexagons; analysis of these LMC data have been presented in \citet{Nidever2019} (hereafter N19). 

Although all fields used in this study reach an unprecedented depth (for SMC studies) of $g \approx 26$ mag, to take on a proper scientific analysis with a reliably extracted final catalogue to such a depth requires several critical steps, including assessing completeness and undertaking careful star-galaxy separation. One way to solve the star-galaxy separation issue is to assess quantitatively the ``goodness of fit'' of the Point Spread Function (PSF) derived in the course of the photometric reduction process.
Given the photometric depth of SMASH, this strategy has the caveat that for saturated stars the pipeline gives a low stellar probability, whereas for very distant and compact galaxies and AGNs found at fainter magnitudes, the pipeline can assign a high stellar probability. 

\begin{figure}
\centering
\includegraphics[width=\columnwidth]{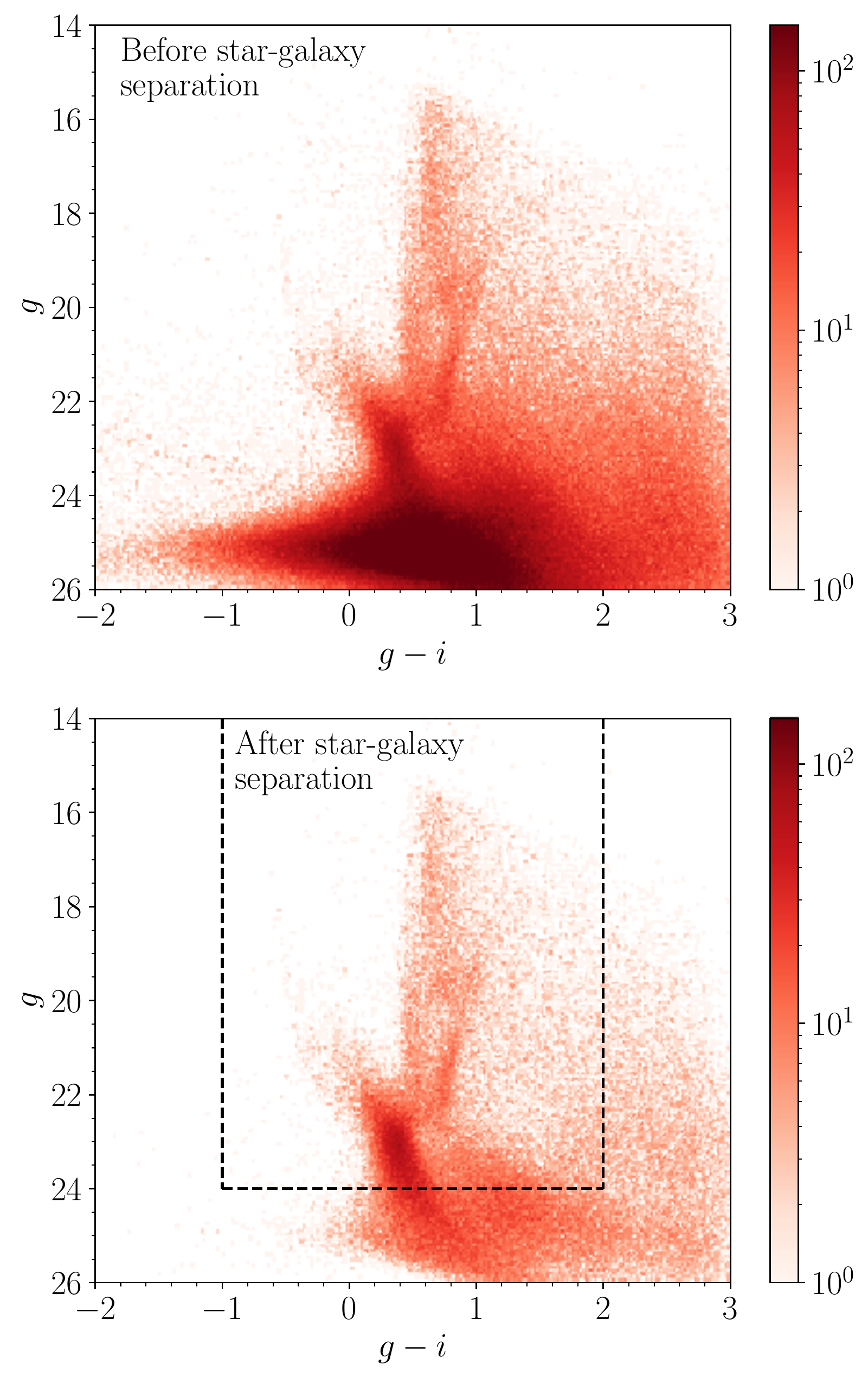}
\caption{Hess diagrams of SMASH field 13, colour-coded by star counts in each bin. The top panel shows the raw diagram before the cuts applied to separate galaxies from stars while the bottom panel shows the clean diagram with the selected stars. (see Section \ref{subsec:data_preparation} for detailed information on the cuts). The dashed area signals the region used for the present study, that avoids MW dominated areas and faint sources with larger errors.}
\label{fig:star-galaxy}
\end{figure}

A more effective approach to tackle this issue is to use, in addition, cuts in colour-colour space, where stars have a well-defined locus in contrast to galaxies. Because SMASH employs several passbands, it gives us the perfect opportunity to explore multiple colour-colour spaces to further refine  classification. Using $g-z$, $r-z$ and $i-z$, with $g-i$ as the fiducial colour, the multi-dimensional stellar locus is computed using bright objects and then a selection envelope defined around that locus is applied to the full sample. For a more detailed explanation on the process of star-galaxy separation we refer the reader to Section 3.1 of N19 who applied the same cuts to SMASH LMC fields. Specifically, their figure 2 shows a visual overview that includes the location of the removed galaxies in a CMD. An example of the result of this process is shown in Figure \ref{fig:star-galaxy}, where we present the CMD of SMASH field 13, located $\sim 5.9$ degrees from the SMC centre (see Figure \ref{fig:study_map}), before applying the colour-colour cuts (top panel) and after the cuts (bottom panel). Some of the predominant SMC stellar population sequences are visible in the ``cleaned'' CMD (bottom panel), such as the MSTO (starting at $g \sim 22$ mag), the red giant branch (RGB), and the RC. In the bottom panel of Figure \ref{fig:star-galaxy}, it is clear that most of the sources that populated the faint part of the CMD (top panel) disappear after applying the cuts. This fact illustrates how insidious the effect of galaxies can be at these magnitudes, whereas all the CMD main features can still be clearly seen.

To avoid any leftover galaxies in our sample, we restrict our analysis to $g \leq 24$ mag. This can be done without significantly affecting our analysis of the SMC surface brightness because most of the integrated light comes from the bright stars, in particular, very young stars as well as giants, lying above the MSTO at $g \approx 22$ mag. Finally, to reduce further contamination we also limit our sample to $-1 \leq g-i \leq 2$ and $g \geq 14$ mag, because no significant or relevant SMC stellar population lies outside of these ranges (see dashed area in the bottom panel of Figure \ref{fig:star-galaxy}). Since there will still be some MW foreground contamination within these limits, we will discuss the procedure to minimise these effects in Section \ref{sec:CMDdecontamination}.

To address completeness, we performed Artificial Star Tests (ASTs), where mock images are created, with a uniformly sampled distribution of colours and magnitudes, and run through the same pipeline as the observed data, including the star-galaxy separation procedure. These are meant to assess the recovery ratio of stars in particular areas of the CMD and are then used to adapt our models to the characteristics of the observations (see Section \ref{subsec:MW_model} for information about the models).

All magnitudes and colours in this study have been reddening corrected using \citet{Schlegel1998} (SFD98) dust maps.  For the innermost parts, we used the value suggested by SFD98 of $E(B-V)=0.037$ mag. We converted the $E(B-V)$ values to $E(g-i)$ using updated bandpass coefficients from \citet{Abbott2018} with $R_V = 3.1$, that reflect the \citet{Schlafly2011} adjustments. Additionally, the SFD98 maps seem to consistently underestimate reddening values for most fields. To address this issue we performed an analysis of the blue edge of the MSTO of the MW halo in both,  the MW theoretical models and the SMASH data. We used the differences in $E(g-i)$, converted to $E(B-V)$, for both blue edges to refine the reddening/absorption corrections. We expect to have SMC reddening maps coming from RC characterisation using SMASH (Choi et. al., in preparation),  similar to those for the LMC \citep{Choi2018a}.
  
\begin{figure}
    \centering
    \includegraphics[width=\columnwidth]{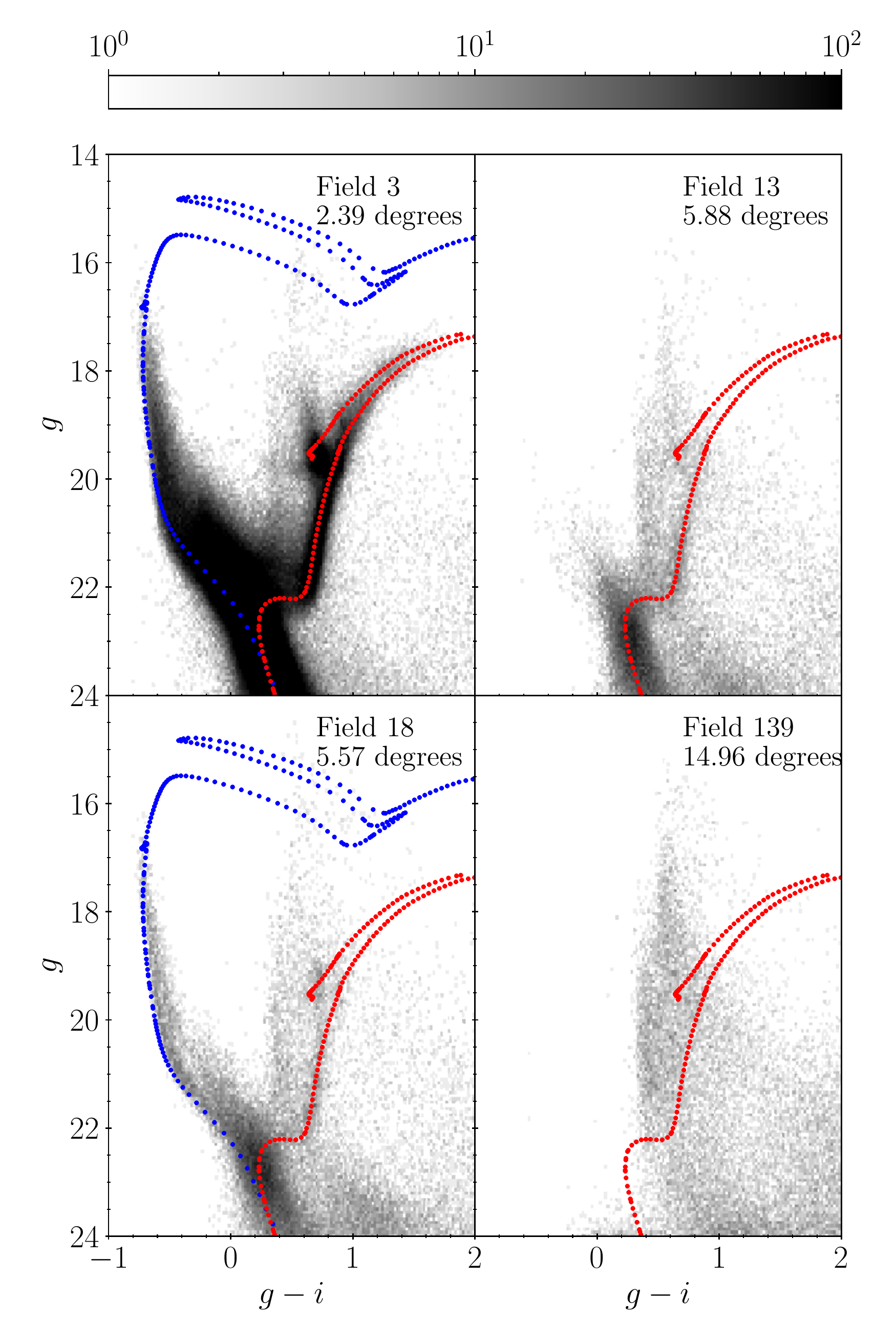}
    \caption{Hess diagrams, saturated at 100 counts, for four selected SMASH fields with over-plotted PARSEC isochrones \citep{Bressan2012}. The red isochrones represent and old stellar population of $\mathrm{[Fe/H]} = -1$ and $\log \mathrm{(yr)} = 9.9$, whilst the blue isochrones represent a younger population with $\mathrm{[Fe/H]} = -0.4$ and $\log \mathrm{(yr)} = 8$. In all cases we used a distance modulus of 18.9. Field 3 is an example of a field in the main body of the SMC that presents both types of populations. Field 13 is in the outskirts of the galaxy with no apparent younger populations. On the contrary, field 18 is inside the Magellanic Bridge, that harbours star forming regions. Field 139 is a typical halo area with no visible SMC stars.}
    \label{fig:example_cmds}
\end{figure}

To make the final selection of fields presented here, we limited it to those with a maximum projected separation of 15 degrees from the SMC centre, represented in green in Figure \ref{fig:study_map}; followed up by a visual inspection of their CMDs. Fields beyond this radius showed no obvious stellar contribution from the SMC and introduced severe contamination from the MW's disc. Figure \ref{fig:example_cmds} shows a sample of CMDs from fields 3, 13, 18, and 139, all located at different radii. Isochrones were over-plotted to indicate the location of stellar populations of the SMC. The red isochrone corresponds to a stellar population with $\mathrm{[Fe/H]} = -1$ and $\log \mathrm{(yr)} = 9.9$ and the blue isochrone depicts a stellar population with $\mathrm{[Fe/H]} = -0.4$ and $\log \mathrm{(yr)} = 8$. We see that field 139, at almost 15 degrees from the SMC centre, shows no contribution from SMC populations. In contrast, we can see SMC stars in field 13, at 5.9 degrees from the centre, matched with an old isochrone. We see young and old stellar populations belonging to the SMC in field 3, located 2.3 degrees from its centre. Field 18, at $\sim 5.6 $ degrees, shows signs of very prominent recent star formation belonging to the Magellanic Bridge, as well as some traces of intermediate stellar population with a visible RC. Finally, field 139 is chosen as a typical halo field and is our furthest field at almost 15 degrees from the SMC centre.

\section{SMC membership} \label{sec:CMDdecontamination}

To avoid the contribution from foreground MW stars, we performed a two-step decontamination process. The first step was to remove the two GCs, 47 Tuc and NGC 362, present in Fields 4 and 9, respectively. The second step was to use  stellar population modelling to remove the foreground contamination from the halo and disc of the MW.

\subsection{47 Tuc} \label{subsec:GC_removal_47 Tuc}

\begin{figure}
\centering
\includegraphics[width=\columnwidth]{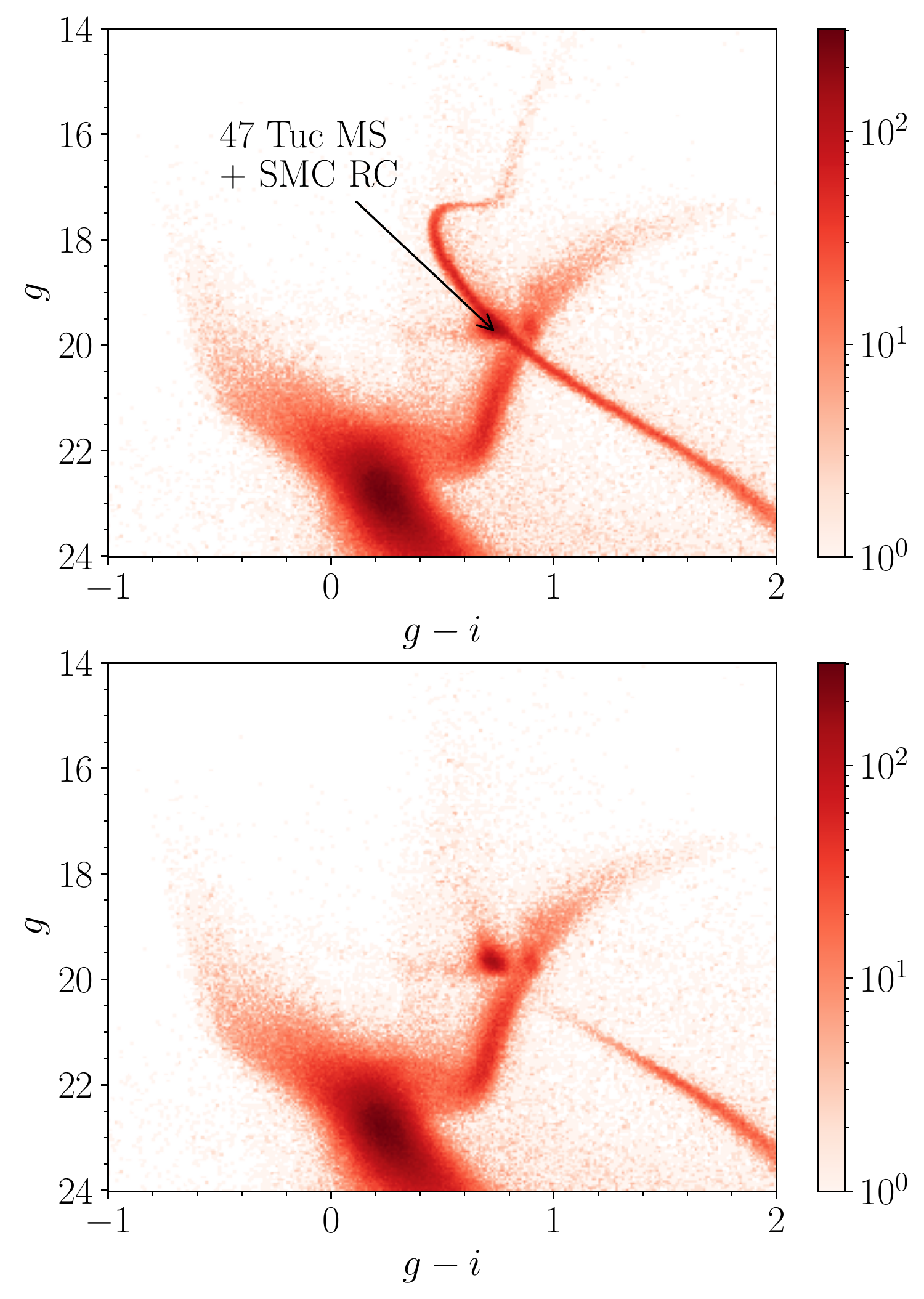}
\caption{Hess diagrams of SMASH Field 4, colour-coded by star counts in every bin. Both show contributions from the SMC and 47 Tuc stellar populations. \textit{Top:} Raw diagram of Field 4, where the MS of the cluster overlaps with the RC population of the SMC. \textit{Bottom:} Cleaned diagram after removing most of the contribution of the cluster using kinematical data from \textit{Gaia} DR2, to $g \sim 20.5$ mag, the limiting magnitude of proper motions.}
\label{fig:47 Tuc_cleaning}
\end{figure}

Because the models we use to remove the MW foreground (Section \ref{subsec:MW_model}) do not account for GCs, their contamination had to be treated separately. Considering that our SMC analysis is based on CMDs, where there is some strong overlap of GC and SMC stars, finding some way to isolate and eliminate the GC stars other than using the CMD is optimal. Figure \ref{fig:47 Tuc_cleaning} shows the CMD of SMASH Field 4 before (top panel) and after (bottom panel) applying the decontamination process. In the top panel we can see that the MS of 47 Tuc overlaps the RC locus of the SMC,  a vital CMD region for population synthesis and brightness calculation.

A simplistic first approach would be to avoid the area in the sky where the GCs are located.  After several tests, we found that to be effective, we had to cut a circular area with a radius of 50 arcmin in Field 4, that translates into $\sim$ 80\% of the total area of Field 4. Hence, we ruled out this highly inefficient approach. 

Fortunately, \textit{Gaia} Data Release 2 \citep[DR2;][]{Gaia2018A} and its accurate proper motion data (with typical uncertainties <1 mas for $G<20$ mag) provides an effective means for discriminating GC stars from SMC stars. 
We can discriminate SMASH stars in these two populations based on their astrometric properties, with the caveat that we can only apply this strategy to stars brighter than $G \approx$ 20.5 mag, the limiting magnitude for the \textit{Gaia} astrometric solution. This results in the lower MS of the cluster still being present in our sample. For 47 Tuc this is not a problem because the region of overlap between the cluster stars and the SMC stars in the CMD is within {\it Gaia} range, $g \sim 20$ and $g- i \sim 0.75$ (see top panel of Figure \ref{fig:47 Tuc_cleaning}). 

To make a robust determination of the kinematical properties of both SMC and 47 Tuc stars, we retrieved a region of 1.5 degrees in radius around the centre of the cluster. This sampling area ensured that almost all cluster stars were included and that there were enough SMC stars to characterise that lower density population in this field. 

To exploit the \textit{Gaia} catalogue fully, we first used its parallax measurements, that for the SMC stars are expected to be lower than 17$\mu$as for an assumed distance of $\sim$60 kpc. This number is lower than the current precision of the catalogue at almost all magnitudes. Therefore we can say that the parallaxes for the SMC stars need to be compatible with having no parallax in this particular dataset. On the other hand, the distance to 47 Tuc is found to be $\sim$4.5 kpc in \textit{Gaia} DR2 \citep{Chen2018}, which corresponds to a parallax of $\sim$0.22 mas. Since we were mainly interested in modelling the SMC and GC stars, we set up thresholds in the parallax to remove some of the other MW contaminants in the sample, because stars in the disc are relatively close and will have large parallaxes. We set an upper limit of 1 mas to our working sample to ensure that we kept all the relevant stars for the study. We then further reduced our sample by doing some proper motion cuts to help with the kinematical analysis. To make sure we took the full distribution whilst removing further MW contaminants, we applied the following cuts in proper motion space, that removed primarily MW halo stars: $-6\, {\rm mas/yr}< \mu_{\alpha} \cos{\delta} < 12\, {\rm mas/yr}$ and $-9\, {\rm mas/yr} < \mu_{\delta} < 4\, {\rm mas/yr}$.

We used two Cauchy distributions to model the kinematical properties of 47 Tuc and the SMC, since they proved to be more successful at reproducing the proper motion distributions (as seen also in \citealt{Chen2018}). These distributions typically are more peaked and have longer tails compared to Gaussian distributions. They are favoured in this case because the uncertainties are not constant inside the sample and increase with the magnitude of the object. The basic equation for the probability density function (PDF) in one dimension is

\begin{equation}
    f(x) = \frac{\gamma}{\pi [(x-x_0)^2 + \gamma^2]},
    \label{eq:cauchy_general}
\end{equation}
where $\gamma$ controls the width of the distribution and $x_0$ its peak value (note that this distribution does not have a mean). 

To determine the parameters of the distributions we used a Markov chain Monte Carlo (MCMC) algorithm, implemented with the Python package \textsc{emcee} \citep{Foreman-Mackey2013}, that uses the following likelihood function:

\begin{multline}
     \mathcal{L}(\mu_{\alpha} ^*, \mu_{\delta}) =  \frac{p \cdot \gamma_{SMC}}{\left[ (\mu_{\alpha} ^* -\mu_{\alpha, \, SMC} ^*)^2 + (\mu_{\delta}-\mu_{\delta, \, SMC})^2 + \gamma_{SMC}^{2} \right] ^{1.5}} + \\
     \frac{(1 - p) \cdot \gamma_{GC}}{\left[ (\mu_{\alpha} ^* -\mu_{\alpha, \, GC} ^*)^2 + (\mu_{\delta}-\mu_{\delta, \, GC})^2 + \gamma_{GC}^{2} \right] ^{1.5}}.
    \label{eq:cauchy_47 Tuc}
\end{multline}
where $p$ represents the fraction of stars belonging to the SMC as compared to the total SMC+GC sample ($0 \leq p \leq 1$), $\gamma_{SMC}$ and $\gamma_{GC}$ are the scale parameters of each distribution and $(\mu_{\alpha, \, SMC}^*, \mu_{\delta, \, SMC})$ and $(\mu_{\alpha, \, GC}^*, \mu_{\delta, \, GC})$ are the proper motions for the SMC and 47 Tuc, respectively. To introduce the observational uncertainties we run the algorithm 200 times using values for the proper motions sample following a 2-dimensional Gaussian distribution characterised by the covariance matrix given in the \textit{Gaia} dataset (columns \texttt{pmra\_error}, \texttt{pmdec\_error}, \texttt{pmra\_pmdec\_corr}). For each of these sets of values, we run 2000 iterations of the sampler and discard the first 400 iterations (burn-in).

To calculate membership probabilities to 47 Tuc we used the probability of a star belonging to the proper motion distribution of the cluster based on the inferred parameters. This is then combined with the probability using an on-sky separation ($\Delta\theta$) to the cluster centre. To simplify things this is taken to be a Gaussian distribution. For this cluster in particular, we took an effective radius ($R_{eff}$) of 20 arcmin. We use equation \ref{eq:membership_47Tuc} that is scaled to have a value of $1$ when a star is at the centre of each of the two distributions,

\begin{equation}
    P_{mem} =  \frac{\gamma_{GC}^3 \exp \left( - \frac{\Delta \theta^2}{2 R_{eff} ^2 } \right)}{\left(\gamma_{GC}^2 + (\mu_{\alpha} ^* -\mu_{\alpha, \, GC} ^*)^2 + (\mu_{\delta}-\mu_{\delta, \, GC})^2\right)^{1.5}}.
    \label{eq:membership_47Tuc}
\end{equation}

Due to the peaked and long-tailed nature of the Cauchy distribution, taking a threshold of $0.5$ for the membership probability was too restrictive. Instead, we based out threshold on an iterative visual inspection of the remaining CMD and proper motion distribution with different thresholds. It was noted that SMC stars had mostly values of membership lower than $0.01$, therefore this value was taken as the discriminant of both populations. These stars in the \textit{Gaia} catalogue were then cross-matched with the SMASH catalogue and removed from further analysis.

\begin{figure}
    \centering
    \includegraphics[width=\columnwidth]{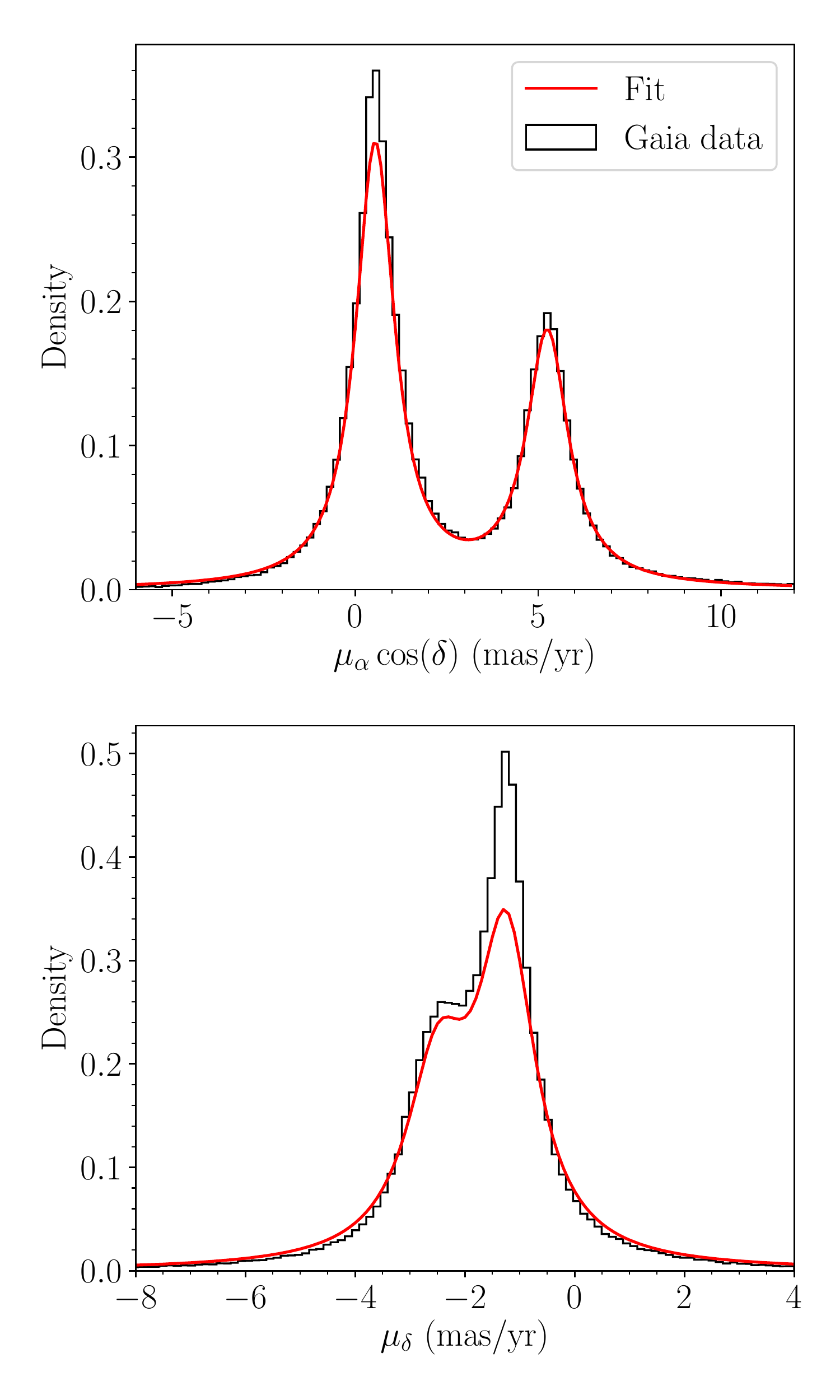}
    \caption{Normalised proper motion distributions of the stars in the field of the cluster 47 Tuc in both right ascension ($\alpha$; top) and declination ($\delta$; bottom). The more populated distribution corresponds to the SMC. In red is shown the fit performed using an MCMC algorithm that searched for two Cauchy distributions in a 2-dimensional space composed by both proper motion components.}
    \label{fig:47_fitting}
\end{figure}

The results of the kinematical separation are summarised in Table \ref{tab:47 Tuc_fit_results} and can be seen in Figure \ref{fig:47_fitting}, where we show the fitted and observed proper motions distribution. The top panel shows the fitting for the right ascension component of the proper motion ($\mu _{\alpha} \cos \delta$) and the bottom one shows the fit for the declination component ($\mu _{\delta}$). Both struggle to match the height of the proper motion peaks, but they do a good job in reproducing their general shape. This is because the fit is normalised for all proper motion space but the data have cuts outside of the area plotted in Figure \ref{fig:47_fitting}.
This is not a problem for our purposes since we only need the median value and have the sole aim of characterising the  width of the distribution. 

The difference between the proper motion of 47 Tuc and that of SMC
stars at the location of 47 Tuc is similar to that reported by
\citet{Bellini2017} from HST data. The proper motion for 47 Tuc is also compatible with \citet{Niederhofer2018a} using the ground-based VMC data. The proper
motion of the SMC stars measured here differs from the proper motion
of the SMC centre of mass as reported by \citet{Kallivayalil2013}. This is due to the internal kinematics of the SMC (for further details see \citealt{DeLeo2020}), as
well as field-dependent viewing perspective. The sense and size of
the field dependence implied by our measurement is qualitatively
consistent with that reported in figure 7 of \citet{Zivick2018}.

While some of the contribution in the faint part of the cluster's MS is still present in the decontaminated sample (see lower panel of Figure \ref{fig:47 Tuc_cleaning}), it does not influence our analysis, as we explain in Section \ref{subsec:MW_model}, where we show the cuts performed to extract the SMC stars from the CMDs.

\begin{table}
    \centering
    \begin{tabular}{c c}
       Parameter  &  Value \\
       \hline
        $p$ & $0.6188 \pm 0.0004$ \\
        $\mu_{\alpha, \, SMC}^{*}$ (mas/yr) & $0.547 \pm 0.002$ \\
        $\mu_{\delta, \, SMC}$ (mas/yr)&  $-1.2453 \pm 0.0013$\\
        $\gamma_{SMC}$ & $0.6406 \pm 0.0011$ \\
        $\mu_{\alpha, \, GC}^{*}$ (mas/yr) & $5.252 \pm 0.002$ \\
        $\mu_{\delta, \, GC}$ (mas/yr) & $-2.494 \pm 0.002$ \\
        $\gamma_{GC}$ & $0.6918 \pm 0.0013$ \\
    \end{tabular}
    \caption{Summary of the values that maximise the likelihood function in Equation \ref{eq:cauchy_47 Tuc} for the SMC and 47 Tuc, applied to \textit{Gaia} DR2 data
    in a radius of 1.5 degrees around the cluster and using an MCMC sampling techinque.}
    \label{tab:47 Tuc_fit_results}
\end{table}

\subsection{NGC 362} \label{subsec:GC_removal_ngc362}

\begin{figure}
    \centering
    \includegraphics[width=0.45\textwidth]{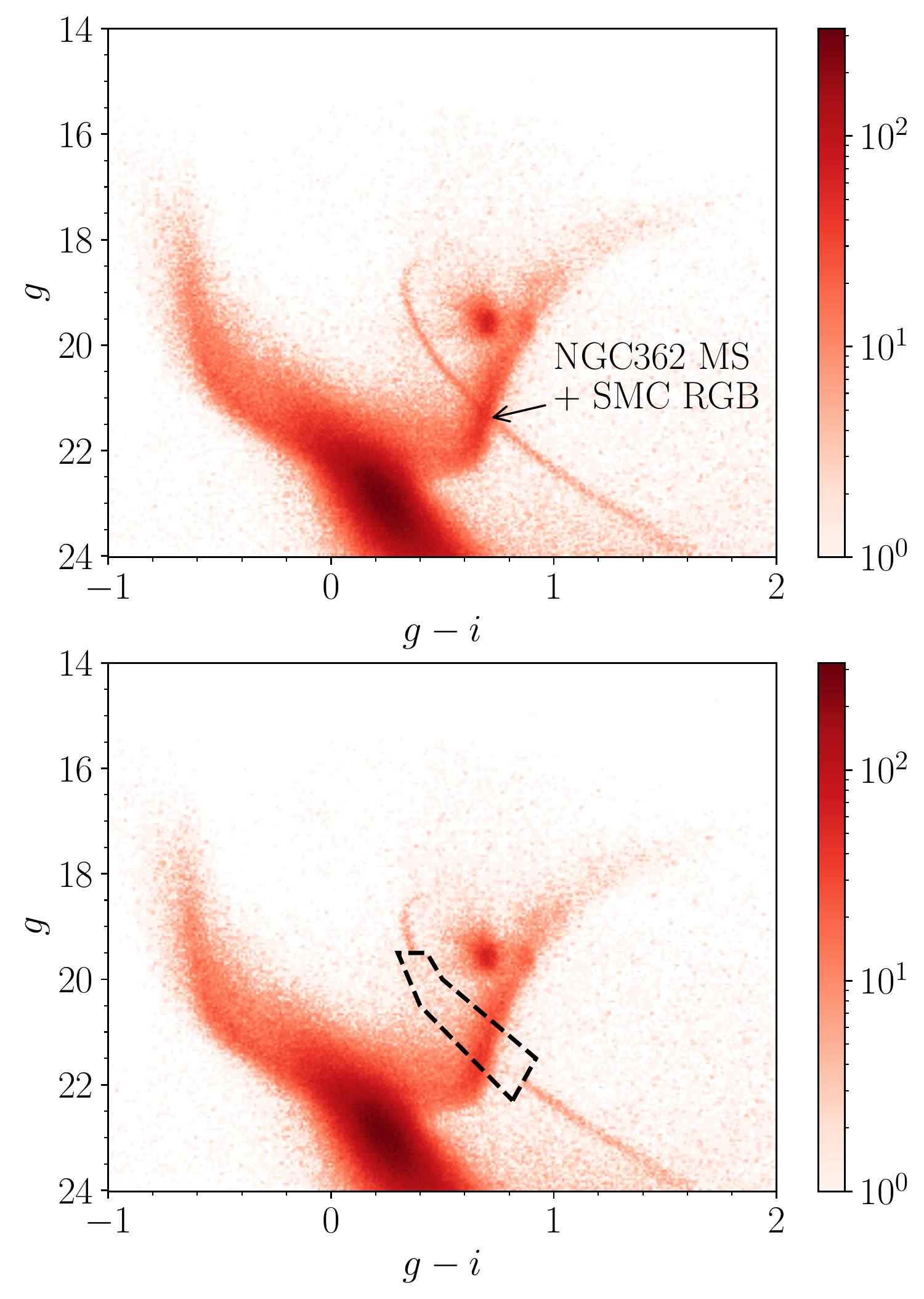}
    \caption{Same as Figure \ref{fig:47 Tuc_cleaning} but for SMASH field 9 and the cluster NGC 362. In this case, the cluster was removed using an isochrone-based approach due to the lack of \textit{Gaia} proper motions at the relevant     magnitudes. The results show a considerable reduction of the GC's population inside the dashed area (bottom panel), also populated by SMC stars.}
    \label{fig:NGC362_cleaning}
\end{figure}

NGC 362 is at a larger heliocentric distance than 47 Tuc, which results in a CMD overlap at a fainter magnitude than the \textit{Gaia} DR2 limit. Therefore, to effectively remove NGC 362 from the area in the CMD occupied with SMC stars, we assigned membership probabilities of the stars in the overlapping regions based on their distance to the MS part of an isochrone and the cluster centre in the sky. The isochrone that was used is not intended to match the exact characteristics of the cluster since we only need a good fit around a specific part of the MS region; therefore, we only needed to adopt values for the distance and metallicity of the cluster that make its MS go through the MS of NGC 362 in our SMASH field. Note also that the decontaminated area is not the full isochrone (see Figure \ref{fig:NGC362_cleaning}), but the part falling inside the SMC area defined in Figure \ref{fig:modelzone_cmd}. To calculate the membership probability we used the following equation 

\begin{equation}
    P_{mem} =  \exp \left( -\frac{\Delta (g-i)^2}{2 \sigma_c ^2 } - \frac{\Delta g^2}{2\sigma_m ^2 } - \frac{\Delta \theta^2}{2 R_{eff} ^2 } \right),
    \label{eq:membership_NGC362}
\end{equation}
where $\Delta (g-i)$ and $\Delta g$ are the differences in colour and magnitude (respectively) between the star and the isochrone, and $\sigma_c$ and $\sigma_m$ are the typical spread in colour and magnitude from the isochrone. These spreads are taken to be $\sigma_c = 0.05$ and $\sigma_m = 0.3$ by visual inspection of the stellar MS of the cluster. Finally $\Delta \theta$ is the angular separation between each star and the centre of the cluster and $R_{eff}$ is the effective radius of the cluster. Based on a visual inspection of the distribution of the stars around the cluster centre, this radius is taken to be $15 \arcmin$.

In Figure \ref{fig:NGC362_cleaning} we present the CMD of SMASH Field 9 before (top panel) and after (bottom panel) the GC contamination. With the adopted classification algorithm we separated most of the GC contribution from the RGB and RC loci of the SMC, as shown in Section \ref{subsec:MW_model}. Nevertheless, due to the statistical nature of the process there is still some contamination present. We take this into account in our analysis by counting the number of stars inside and outside of a small area around the isochrone to make an estimate of the potential error in the flux that it will yield.

\subsection{Milky Way Modelling} \label{subsec:MW_model}

To clean the observed CMDs of MW contamination, we used the MW models developed by de Boer et al. (in preparation). The use of custom made MW models is necessary because popular models such as the Besan\c{c}on Galactic Model \citep{Robin2003} or Galaxia \citep{Sharma2011} provide unsatisfactory results for CMDs down to the depth probed by SMASH. In particular, the MW halo populations at these magnitudes are not well reproduced, even though these are the main contaminants in the MSTO region of the LMC/SMC. 

The custom models were constructed by fitting a set of MW components with parameters similar to those of Galaxia \citep{Sharma2011}, with age and metallicity parameters (including Gaussian widths) left free.
The models were fitted to a subselection (38 fields) of SMASH fields, that were chosen to be far enough away from the MW midplane ($\mid$ b $\mid$ $>$20 deg) and the MCs ($>$10 deg from each) in order to avoid populations not included in the model. The models were constructed using the MESA isochrones and stellar tracks (MIST) models \citep{Choi2016} in the region $-$0.2$<$$g-i$$<$1.8 and 14$<$$g$$<$22.5 to sample CMD regions that exclude faint M-dwarfs (that are challenging to model) and where the data are 100\% complete, as shown in N19. The resulting best-fit parameters for the thin disc, thick disc and halo are mostly consistent with previous studies. Of most relevance to the present work, the best fit to the data gives a MW halo density law that follows a power law index of $-$3 and flattening $\sim$0.5, similar to those seen previously,  important for reproducing the faint halo component \citep{Bell2008,Deason2011,Slater2016}. With parameters in place, MW models were generated for each SMASH field using the best-fit parameters, and convolved with the photometric errors and completeness (see Figures~2 and~3 in  \citealt{Nidever2019} for more details about completeness) of each individual field, so that the model distributions could be directly subtracted from the observed CMDs.

The models were created for nominal 3 deg$^2$ DECam fields; this is, however, a coarse approximation in SMASH, given that some of the CCDs of DECam are not included in some fields. We corrected this by calculating the real observed area for every field, creating a grid over the sky with a binning that depends on the stellar density of each field. This is to prevent not accounting for areas that do not have any stars in them but have been observed. We then counted the number of bins with stars and summed their areas; this introduces some uncertainty that will be discussed in Section \ref{subsec:errors}.

For a further sanity check we analysed the fields that are far enough from the MCs that only have MW contribution. The resultant number of remaining stars after the subtraction process was found to be exponentially dependent on the Galactic latitude ($b$). To correct for this, we scaled the number of counts in each mock catalogue based on the relation between the residual counts of these outer fields and $b$. We fitted an exponential relation and extrapolated it to the SMC fields. After this, the remaining foreground contamination was more homogeneous and compatible with gaussian-like uncertainties. 

\begin{figure}
    \centering
    \includegraphics[width=\columnwidth]{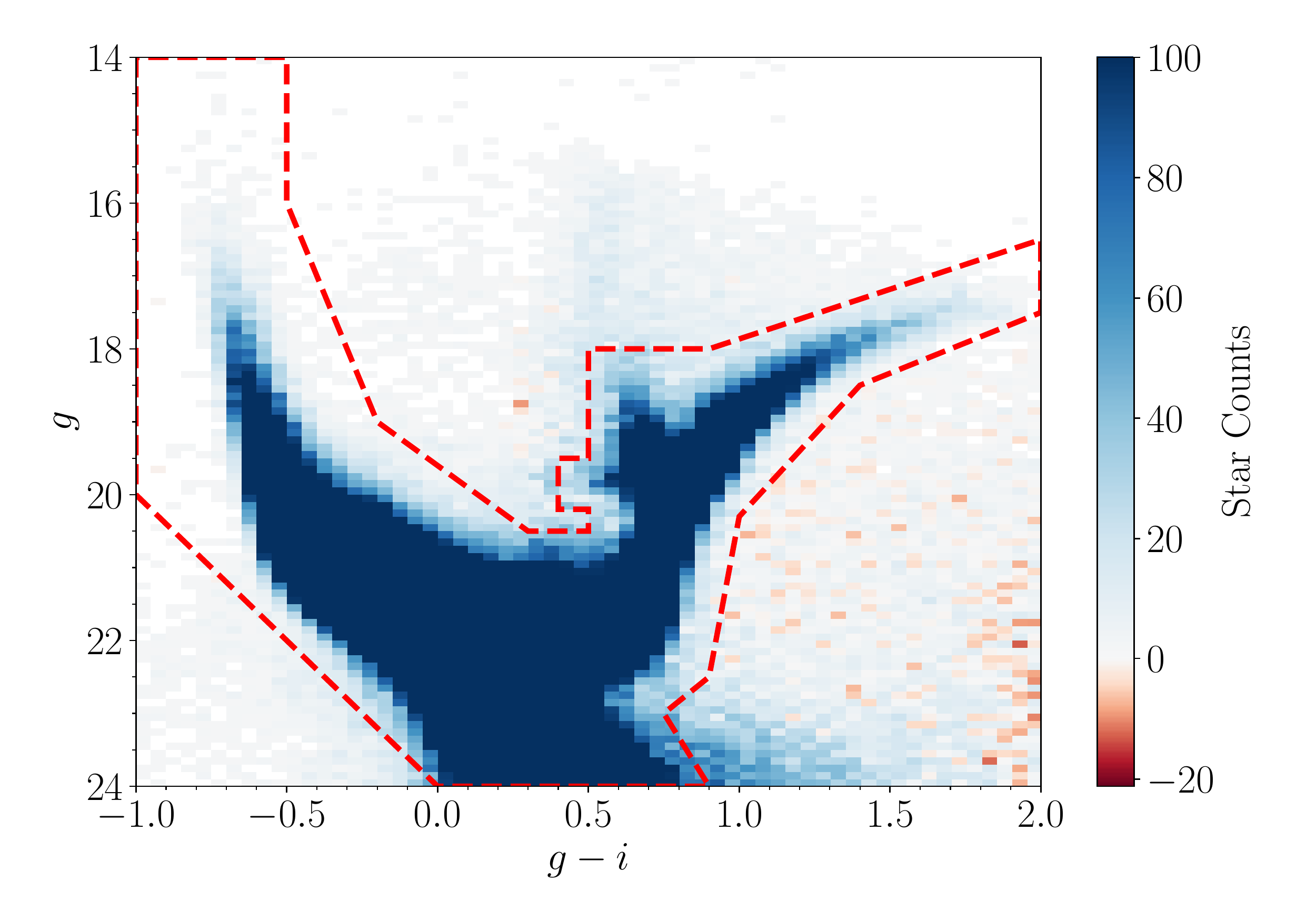}
    \caption{Hess diagram of SMASH field 3 after the subtraction of foreground stars using MW modelling. The binning of the data corresponds to the actual binning used to make the analysis, that allows for an appropriate subtraction of the model. The red dashed line is the border of the area from where we take the SMC stars (see Section \ref{subsec:MW_model} for details on the foreground subtraction). The histogram is coloured according to the star count in each bin, as shown in the colour bar.}
    \label{fig:modelzone_cmd}
\end{figure}

Finally, to ensure that we get the most out of our data, we used cuts in the CMDs of each field to further optimise the selection of SMC stars. An example of the cuts applied for each field can be seen in Figure \ref{fig:modelzone_cmd}, where we present a Hess diagram of SMASH Field 3 after the subtraction of foreground stars using MW modelling. The area contained by the red dashed polygon, which remains unchanged from field to field, indicates the region of stars that has been used to calculate the brightness of the SMC and the background histogram shows the number of stars in each bin. The sum of the flux of all these stars represents our final calculation of the brightness of each field. In this example, we have taken a central field of the SMC that has a representation of all the different stellar populations present in the diagram, ranging from young stars to RGB and RC stars, as well as very faint MS stars. With this selection we were also able to remove the rest of the contribution from the globular clusters in Fields 4 and 9.

\subsection{Assessment of uncertainties} \label{subsec:errors}

To derive the surface brightness profile of the SMC we must first convert the  magnitudes of SMC stars in a region as measured by SMASH into a total normalised flux, $F'$, calculated from the $g$ magnitude as $10^{-0.4 g}$. Then we divide the flux by the area, $A$ of the region (which has its own intrinsic error) to convert it into a surface brightness, $\Sigma$:

\begin{equation}
    \Sigma = -2.5 \log \left( \frac{F'}{A} \right).
    \label{eq:SB}
\end{equation}

\begin{figure}
    \centering
    \includegraphics[width=\columnwidth]{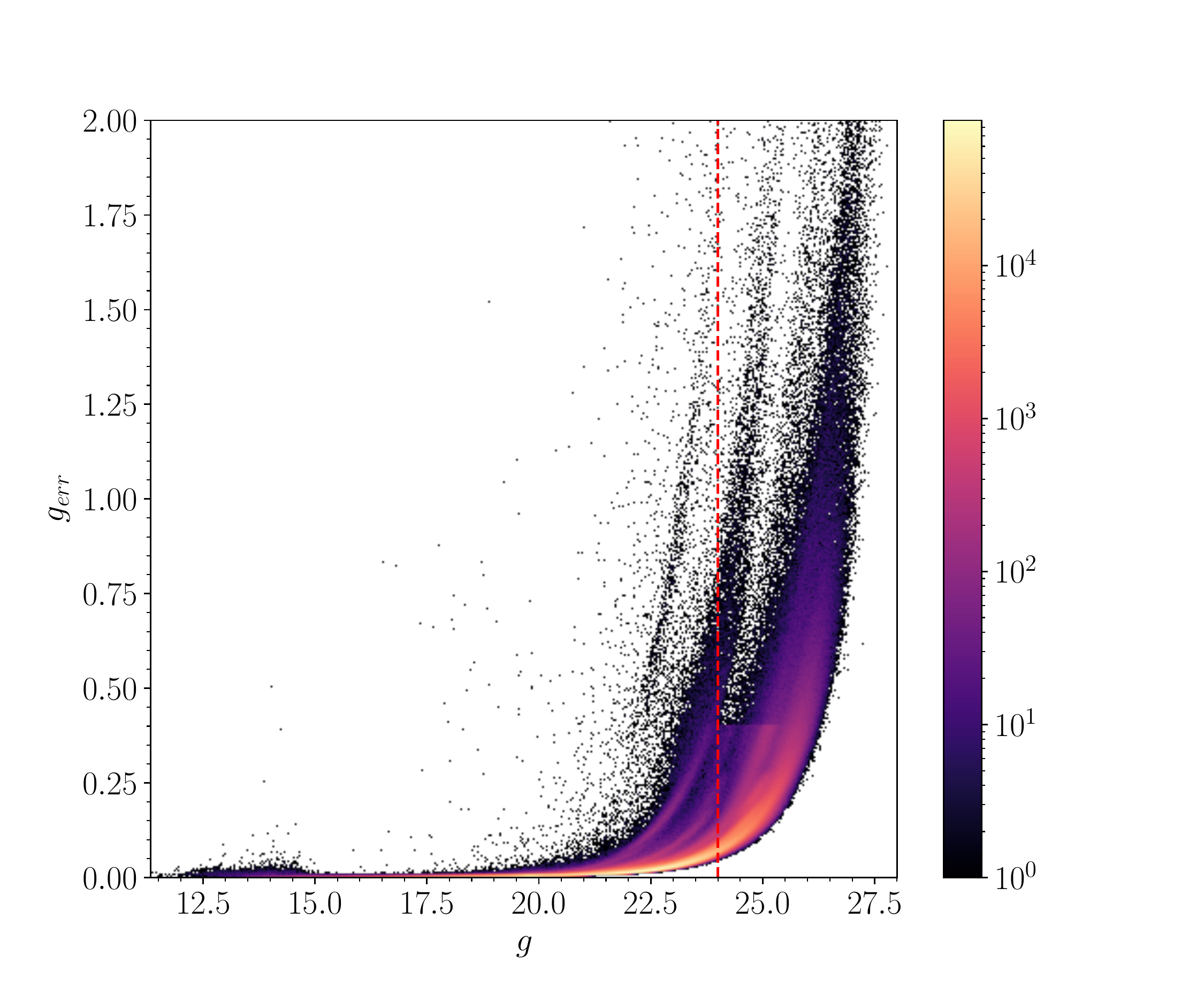}
    \caption{Evolution of the errors in $g$ with magnitude for all fields of SMASH considered in this paper. The red dashed line shows our selection cut for the analysis, which ensures that we have the best quality data. The different trends that are seen correspond to the various fields, that have different exposure times and reductions. Field 1, being the shallowest, produces the trend that is seen to increase at brighter magnitudes than the others.}
    \label{fig:g_uncertainties}
\end{figure}

However, among the most crucial aspects of calculating the surface brightness profile of the SMC, particularly at its faintest radii, is to address the levels of both contamination and incompleteness in our SMC sample as discussed in N19. In this subsection we discuss these and other potential sources of error in our measurement of the surface brightness (Equation \ref{eq:SB}). Among them are:

\begin{itemize}
    \item \textbf{Milky Way modelling:} The MW contamination is the most important source of uncertainty and to account for it, we needed to quantify how well the model works. For this purpose, we analysed the subtraction of the MW foreground population in fields well beyond the area of the MCs, where we do not expect any other stellar population besides that from the MW. We found that in all cases we still had foreground stars after the subtraction and that the number of remaining stars is dependant on the Galactic latitude, because of the MW disc. This effect is more prominent in brighter magnitudes ($g<20$), so it means that our results based on the selection of Figure \ref{fig:modelzone_cmd} might be affected by a certain gradient due to the MW disc. To account for that we focused on further homogenising the MW model (more details in Section \ref{sec:SB profile}) and characterising the remaining foreground stars. Our selection sample shown in Figure \ref{fig:msto_cmd_modelzone} is not affected by MW disc and the procedure of characterising the goodness of the MW model will be described in Section \ref{sec: MSTO profile}. 
    \item \textbf{Area uncertainty:} Because some individual CCD detectors for DECam were not functioning during some observations, and because there are some excluded areas for each field for various reasons (e.g. due to extremely bright stars, saturated cluster cores, or crowding issues, among others), it was necessary to calculate the areas for each field independently. To do so, we subdivided the field into a grid of bins in sky coordinates, we calculated how many bins have objects in them, and finally we summed the area of all of these bins. We then scaled the size of the binning depending on the crowding of the fields. From this analysis we found that the area determination had an associated uncertainty of $\sim 4$\% of the total area of each field, that was then taken into account when we calculated the surface brightness.
    \item \textbf{Observational uncertainties:} In SMASH, special care has been taken to get the best possible determination of the photometry, with a reduction process that involves very precise PSF fitting and very accurate calibration of the different filters using \textit{Gaia} DR2 and DES. For the whole survey, the $5\sigma$ point source depths for $g$ and $i$ are 24.8 and 24.2, respectively. Restricting our sample to $g \leq 24$ ensures that we use the best quality data. In Figure~\ref{fig:g_uncertainties} we see the evolution of the uncertainty of $g$ with magnitude, showing that most of the data for the chosen range (indicated by the red dashed line) has very low uncertainty. The only exception is field 1, which produces the trend that increases at brighter magnitudes due to shallower exposures. These uncertainties though, are not the main concern for the surface brightness profile described in Section \ref{sec:SB profile}, because it is dominated by brighter stars that have very small errors. For the profile shown in Section \ref{sec: MSTO profile}, the effect could be bigger, but our selection (Figure \ref{fig:msto_cmd_modelzone}) ensures that for almost all the survey (except field 1), 99.5\% of stars have errors that are lower than $0.05$ mag proving this is not the dominant effect. For field 1, we find that 98\% of stars have errors in the range $0.05-0.15$. 
    \item \textbf{Milky Way globular clusters:} Every statistically based selection method applied to the data will introduce some noise in the results. Because we used two different methods to remove the contribution from the two globular clusters and the methods have intrinsically  different properties, their impacts on the surface brightness calculated for each respective field will also vary.
    Moreover, the cores of globular clusters in general are very crowded and unresolved with ground-based telescopes. For the two relevant clusters here, the central areas are not resolved and are saturated with light, and this reduces the effective survey area of the field.  We will discuss how this affects each of the analysed clusters.
    \begin{itemize}
        \item \textbf{47 Tuc} is the largest of the two clusters and also the closest to us. These characteristics give us the opportunity to use \textit{Gaia} DR2 to more effectively clean this cluster's stars from our data. On the other hand, the cluster's proximity also produces a bigger sky area with crowding issues. Nevertheless, the reduction in the area due to the crowding at the core is less than 0.1 deg$^2$ which is similar to the area reduction in other fields due to crowding and missing chips.
        \item \textbf{NGC 362:} In this particular case it is clear (Fig. \ref{fig:NGC362_cleaning}) that we still have some contamination from the cluster in our final selection. Given that the field itself is very well populated with SMC stars, the overall effects will be very small, with an estimated $\sim 100$ stars from the cluster in our final sample. This translates into an overestimation of the flux by $\sim$0.02\%. In addition, compared to 47 Tuc, we find an even smaller reduction of the area due to the crowding in the core of less than 0.05 degrees$^2$.
    \end{itemize}
\end{itemize}

In summary, to estimate the uncertainty in the surface brightness, we calculated the uncertainty in both the flux and the area. Propagating the errors of the quantities discussed in Equation \ref{eq:SB}, we obtain

\begin{equation}
    \delta \Sigma = \frac{2.5}{\ln (10)} \sqrt{\frac{\delta F'^2}{F'^2} + \frac{\delta A^2}{A^2}}.
    \label{eq:error}
\end{equation}

\noindent This equation implies that if we assume constant uncertainties for the remaining MW flux and the area across all fields, the error will increase if either the flux or the area decreases. This is a reasonable assumption to make since the error in flux we are considering here is the one coming from the MW subtraction, and we have already secured that it will be constant across fields by scaling according to the Galactic latitude.

\section{Surface Brightness Profile} \label{sec:SB profile}

With a sample of SMC stars cleaned of foreground contaminants according to the methods previously described, we calculated the surface brightness of the SMC for each field.  Note that, despite the fact that in the central regions of the SMC the DECam fields overlap (see Figure \ref{fig:study_map}), we do not account for that overlap in any way. This is because every DECam pointing is treated as an independent surface brightness profile measurement, that has been calibrated individually and separate from the others. It is also not our main priority to have a very precise profile for the main body, as we are more interested in characterising the shape of the profile in the SMC outskirts. 

\begin{figure}
\centering
\includegraphics[width=\columnwidth]{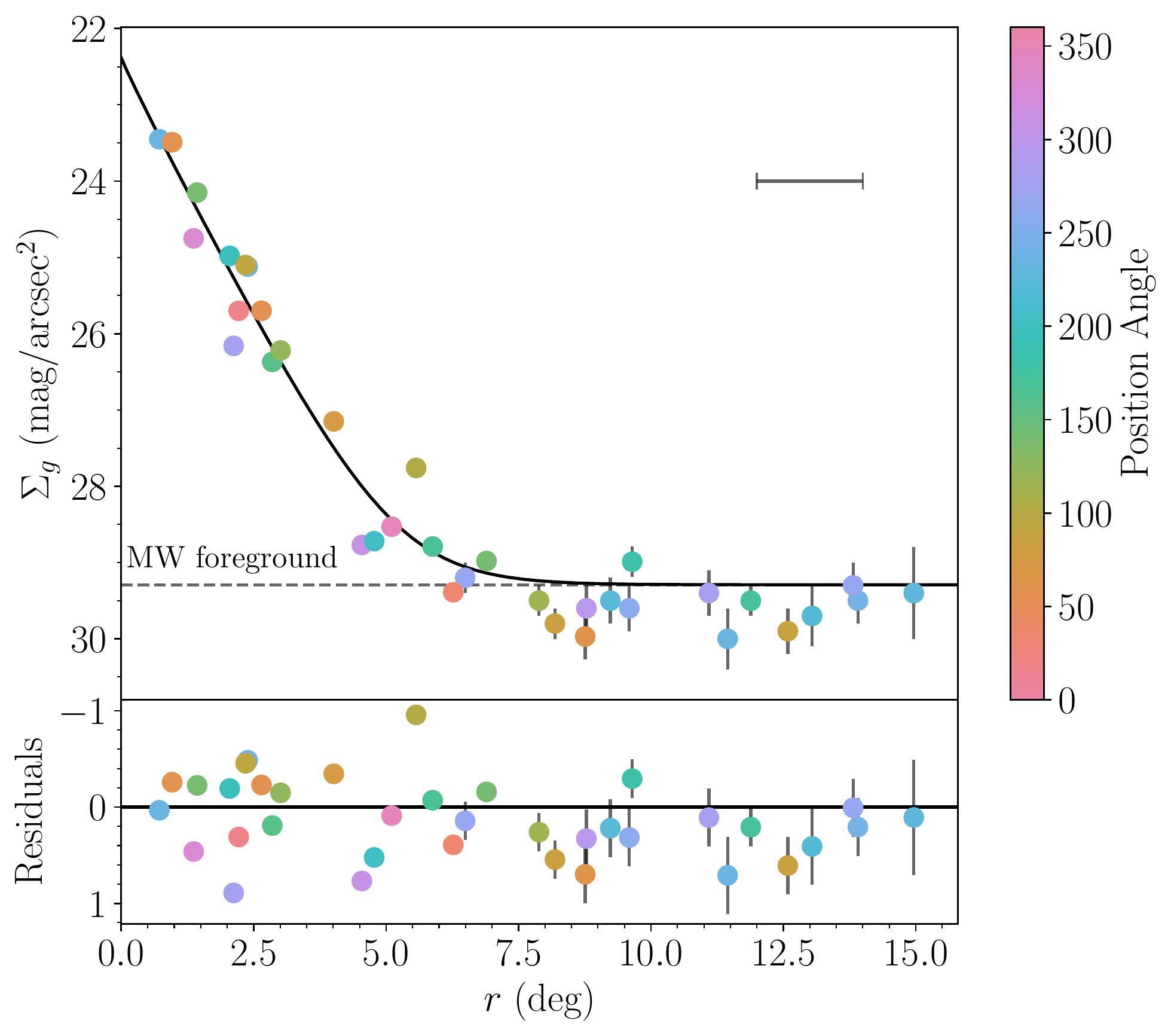}
\caption{Surface brightness profile of the SMC up to 15 deg from its centre. The standard convention for Position Angle (PA) in equatorial coordinates is used: $0\degree / 360\degree$ is North, $90 \degree$ is East, $180 \degree$ is South and $270 \degree$ is West. Each point represents the centre of the SMASH fields. At the same time, each has a radial extent of $\pm 1$ degree from its centre, as represented by the horizontal error bar in the top-right. The LMC in this coordinate system lies in the North-East direction (PA$\sim 100-150$). The black line shows a fit of a S\'{e}rsic profile described in Equation \ref{eq:sersic_profile} with the parameters in Table \ref{tab:profile_fit}. The foreground is reached at $\sim 8$ degrees from the centre of the SMC in these circular polar coordinates.}
\label{fig:SB_radial_profile}
\end{figure}

\begin{table*}
    \centering
    \caption{Data used to construct Figures \ref{fig:SB_radial_profile} and \ref{fig:MSTO_profile_belokurov}. The field numbers (corresponding to the ones assigned in SMASH as displayed in  Figure \ref{fig:study_map}) are shown in column 1; the radial distance from the SMC centre \citep{Paturel2003} and PA of each field are shown in columns 2 and 3, respectively, following the same convention as Figure \ref{fig:SB_radial_profile}. Column 4 presents the surface brightness (in mag/arcsec$^2$) for each field and the estimated uncertainty, all of which calculated using the method explained in Section \ref{sec:CMDdecontamination}. Columns 5 and 6 show the MSTO stellar density and its inferred surface brightness, as explained in Section \ref{sec: MSTO profile}.}
    \begin{tabular}{|l c c c c c |}
    \hline
    Field & r ($\degree$) & PA($\degree$) & $\Sigma_g$ (mag/arcsec$^2$) & MSTO density (stars/$\deg^2$) & $\Sigma_{g, \mathrm{MSTO}}$ (mag/arcsec$^2$)$^a$\\
    \hline
    1 & 4.54 & 306.66 & $28.77 \pm 0.05$ & $2297.1 \pm 102.0$ &  $29.46 \pm 0.05$\\
    2 & 4.78 & 203.22 & $28.72 \pm 0.05$ & $2477.5 \pm 103.0$ & $29.36 \pm 0.05$ \\
    3 & 2.39 & 238.01 & $25.12 \pm 0.04$ & - & - \\
    4 & 2.12 & 278.71 & $26.16 \pm 0.04$ & - & - \\
    5 & 1.37 & 330.22 & $24.75 \pm 0.04$ & - & - \\
    6 & 0.72 & 235.42 & $23.45 \pm 0.04$ & - & - \\
    7 & 2.05 & 194.23 & $24.98 \pm 0.04$ & - & - \\ 
    8 & 9.65 & 180.42 & $28.99 \pm 0.2$ & $272.0 \pm 22.2$ & $31.91 \pm 0.09$ \\
    9 & 2.22 & 18.95 & $25.70 \pm 0.04$ &  - & - \\
    10 & 0.96 & 59.07 & $23.49 \pm 0.04$ &  - & - \\
    11 & 1.43 & 139.54 & $24.15 \pm 0.04$ &  - & - \\
    12 & 2.85 & 158.77 & $26.37 \pm 0.04$ & - & - \\
    13 & 5.88 & 167.86 & $28.79 \pm 0.05$ & $2538.2 \pm 99.7$ & $29.40 \pm 0.04$\\ 
    14 & 2.65 & 57.03 & $25.70 \pm 0.04$ & - & - \\
    15 & 2.35 & 93.72 & $25.10 \pm 0.04$ & - & - \\
    16 & 3.01 & 125.07 & $26.22 \pm 0.04$ & - & - \\ 
    17 & 11.89 & 172.08 & $29.5 \pm 0.2$ & $293.1 \pm 19.4$ & $31.92 \pm 0.07$\\
    18 & 5.57 & 103.66 & $27.76 \pm 0.05$ & $2561.2 \pm 104.9$ & $29.31 \pm 0.04$ \\
    19 & 6.90 & 141.73 & $28.98 \pm 0.06$ & $1216.6 \pm 56.2$ & $30.20 \pm 0.05$\\ 
    20 & 8.76 & 62.40 & $29.97 \pm 0.3$ & $93.4 \pm 16.2$ & $34.03 \pm 0.19$ \\
    21 & 8.19 & 84.12 & $29.8 \pm 0.2$ & $197.1 \pm 21.3$ & $32.29 \pm 0.12$ \\
    22 & 7.89 & 115.44 & $29.5 \pm 0.2$ & $356.1 \pm 23.2$ & $31.61 \pm 0.07$ \\
    23 & 12.58 & 84.52 & $29.9 \pm 0.3$ & $79.4 \pm 16.6$ & $33.6 \pm 0.2$ \\
    139 & 14.96 & 227.34 & $29.4 \pm 0.6$ & $-3.2 \pm 18.3$ & No SMC \\
    141 & 13.04 & 217.13 & $29.7 \pm 0.4$ & $-50.1 \pm 17.1$ & No SMC \\
    142 & 13.91 & 243.87 & $29.5 \pm 0.3$ & $85.2 \pm 14.1$ & $35.17 \pm 0.18$ \\ 
    143 & 11.45 & 235.53 & $30.0 \pm 0.4$ & $9.1 \pm 17.1$ & No SMC \\
    144 & 13.82 & 268.52 & $29.3 \pm 0.3$ & $75.2 \pm 12.7$ & $35.34 \pm 0.18$ \\
    145 & 9.59 & 259.21 & $29.6 \pm 0.3$ & $25.3 \pm 12.1$ & No SMC \\
    147 & 11.10 & 282.54 & $29.4 \pm 0.3$ & $1.5 \pm 25.7$ & No SMC \\ 
    149 & 6.49 & 266.18 & $29.2 \pm 0.2$ & $270.3 \pm 16.6$ & $32.12 \pm 0.07$ \\
    150 & 8.78 & 295.04 & $29.6 \pm 0.3$ & $106.2 \pm 16.4$ & $33.16 \pm 0.17$ \\
    176 & 5.11 & 349.18 & $28.53 \pm 0.04$ & $5716.8 \pm 210.1$ & $28.48 \pm 0.04$ \\ 
    177 & 6.27 & 34.57 & $29.39 \pm 0.06$ & $1391.0 \pm 55.5$ & $30.03 \pm 0.04$ \\
    178 & 4.01 & 72.14 & $27.15 \pm 0.04$ & $15609.9 \pm 550.0$ & $27.36 \pm 0.04$ \\
    181 & 9.23 & 223.67 & $29.5 \pm 0.3$ & $-60.2 \pm 24.5$ & No SMC \\ 
    \hline
    \end{tabular}
    \\
    \vspace{0.1in}
    \footnotesize{$^a$ See Section \ref{sec: MSTO profile} for details.}
    \label{tab:SB_profile}
\end{table*}

The results of our calculations can be seen in the surface brightness profile in Figure \ref{fig:SB_radial_profile}, with the data from each individual field summarised in Table \ref{tab:SB_profile}. Figure \ref{fig:SB_radial_profile} shows the surface brightness profile up to 15 degrees from the SMC centre, with each filled circle representing an individual SMASH field. The colours of the symbols denote the different position angles (PA), as indicated by the scale bar on the right side of the figure. The error bars are small due to the small statistical uncertainty coming from the fact we have a large number of stars (specially in the innermost parts) and that we are averaging over a large area in the sky ($\sim 3 \deg ^2$) to calculate something at the arcsec$^2$ level. It is important to note that the limit of brightness reached in this study is not the one associated with the faintest structures of the SMC. Rather, this is a limit imposed by the methodology followed. Given the limitations of the MW model in the area of SMC RGB stars, and the fact that stars in this area are amongst the brightest of the sample, this method produces a ``noisy'' profile, creating an obstacle to disentangle the very faint structures that might be present (see Section \ref{sec: MSTO profile}). 

As seen in the profile, we detect clear SMC brightness features, in every direction in PA, as far out as 6 degrees. After that, the limits where the surface brightness of the SMC cannot be distinguished from noise introduced by the foreground lie between 6 and 8 degrees depending on the PA. In the outskirts of the galaxy the measured values of the profile seem more staggered. This is due to a combination of the use of a circular polar coordinate system, instead of an elliptical one, and the difficulties in the RGB area to discern MW from SMC. Assuming a distance to the centre of the SMC of 63.4 kpc \citep{Ripepi2016}, this means that there is significant SMC material as far as $\sim 8$ kpc away from its centre. 

We once again employed an MCMC sampling technique (using the same software implementation described in Section \ref{subsec:GC_removal_47 Tuc}), running 5000 iterations (burn-in of 2000) and with 4 free parameters, to obtain a fit to our radial profile using the following S\'{e}rsic profile

\begin{equation}
    \Sigma_g (r) = -2.5 \log \left\{ F_0 \exp\left[ -\left( \frac{r}{h} \right)^{\frac{1}{n}} \right] + F_f \right\}
    \label{eq:sersic_profile}
\end{equation}
where the free parameters are the central flux ($F_0$), the scale-length ($h$), the S\'{e}rsic index ($n$) and the foreground flux ($F_f$). The result is also shown in Figure \ref{fig:SB_radial_profile} as a black solid line, where we have included a residual plot that exemplifies how far the points are from the fit. Table \ref{tab:profile_fit} summarises the derived parameters resulting from the fitting procedure. The behaviour of the fit is generally correct. We note though that the fit is overestimating the foreground level slightly. This is due to the fact that the points located in the outskirts (which should set this value, for the most part), have much bigger uncertainties than the points located in the inner parts. This causes the sampling technique to weight them less. It also points out to the fact that this might be a too simple approximation for a galaxy that is heavily disrupted. For example, field 18, which is on the Magellanic Bridge, has a much brighter surface brightness than the fields in similar radii. Because the uncertainty in this field is smaller, this pushes the foreground to slightly brighter magnitude ($\sim 0.1$ mag). It is important to point out that although the derived index $n = 1.05$ is consistent to that of a disc-like galaxy, this does not imply that the SMC had a disc to begin with, given the big scatter in the profile.

\begin{table}
    \centering
    \begin{tabular}{c c}
       Parameter  &  Value \\
       \hline
        $F_0$ & $(1.11 \pm 0.07) \times 10^{-9}$ \\
        $\Sigma_0$ (mag/arcsec$^2$) & $22.39 \pm 0.07$ \\
        $h$ (degrees) & $0.75 \pm 0.04$ \\
        $n$ &  $1.05 \pm 0.03$\\
        $F_f$ & $(1.92 \pm 0.07) \times 10^{-12}$ \\
        $\Sigma_f$ (mag/arcsec$^2$) & $29.29 \pm 0.04$ \\
    \end{tabular}
    \caption{Summary of the values for the free parameters that fit the function in Equation \ref{eq:sersic_profile}. This can be seen in the SMC surface brightness profile out to a radius of 15 degrees using circular polar coordinates (Figure \ref{fig:SB_radial_profile}). }
    \label{tab:profile_fit}
\end{table}

Using RGB stars, \citet{Nidever2011} found an extended profile for the SMC reaching up to $\sim 12$ degrees from its centre. It is not reasonable to make comparison between our full surface brightness profile and theirs, due to the difference in the methods used. Nevertheless, we took their approach of an elliptic profile to study the distribution in the inner parts, and that probed fairly well by our profile. We used an MCMC algorithm to calculate the semi-major axis of the ellipse that goes through the centre of each field according to the following transformation:

\begin{equation}
    a = \sqrt{ x^2 + \frac{y^2}{(1 - e)^2} }.
    \label{eq:elliptical_radius}
\end{equation}

\noindent where, $x$ and $y$ are Cartesian coordinates whose origin is the centre of the SMC, and $e$ is the eccentricity of the ellipse. Using the corresponding values of $a$ for each of the field radii and embedding the fitting of $e$ and the PA ($\phi_0$) into the S\'{e}rsic profile, we obtained slightly different values than from our circular fitting. These are summarised in Table \ref{tab:profile_fit_elliptical}, where we found an eccentricity for the SMC of $0.155 \pm 0.006$, and a line-of-nodes of $\phi_0 = 73.5 \pm 1.1$ degrees. We found that using elliptical coordinates, the quality of our fits did not improve. This likely means that the dispersion of the points in Figure \ref{fig:SB_radial_profile} comes from both the intrinsically irregular shape of the SMC and the stochastic nature of our method.

\begin{table}
    \centering
    \begin{tabular}{c c}
       Parameter  &  Value \\
       \hline
        $F_0$ & $(1.08 \pm 0.07) \times 10^{-9}$ \\
        $\Sigma_0$ (mag/arcsec$^2$) & $22.42 \pm 0.07$ \\
        $h$ (degrees) & $0.78 \pm 0.04$ \\
        $n$ &  $1.04 \pm 0.03$\\
        $F_f$ & $(2.07 \pm 0.07) \times 10^{-12}$ \\
        $\Sigma_f$ (mag/arcsec$^2$) & $29.21 \pm 0.04$ \\
        $e$ & $0.155 \pm 0.006$ \\
        $\phi_0$ (degrees) & $73.5 \pm 1.1$ \\
    \end{tabular}
    \caption{Summary of the values for the free parameters that fit the function in Equation \ref{eq:sersic_profile}, together with the values for the eccentricity ($e$) and the line-of-nodes of the galaxy ($\phi_0$).}
    \label{tab:profile_fit_elliptical}
\end{table}


\section{Stellar density profile} \label{sec: MSTO profile}

N19 used SMASH data to trace the very faint outskirts of the LMC with stars in the MSTO. In this Section, we perform a similar analysis for the SMC in order to compare the results for both galaxies, and to fairly assess the low surface brightness features around both MCs.

\begin{figure}
    \centering
    \includegraphics[width=\columnwidth]{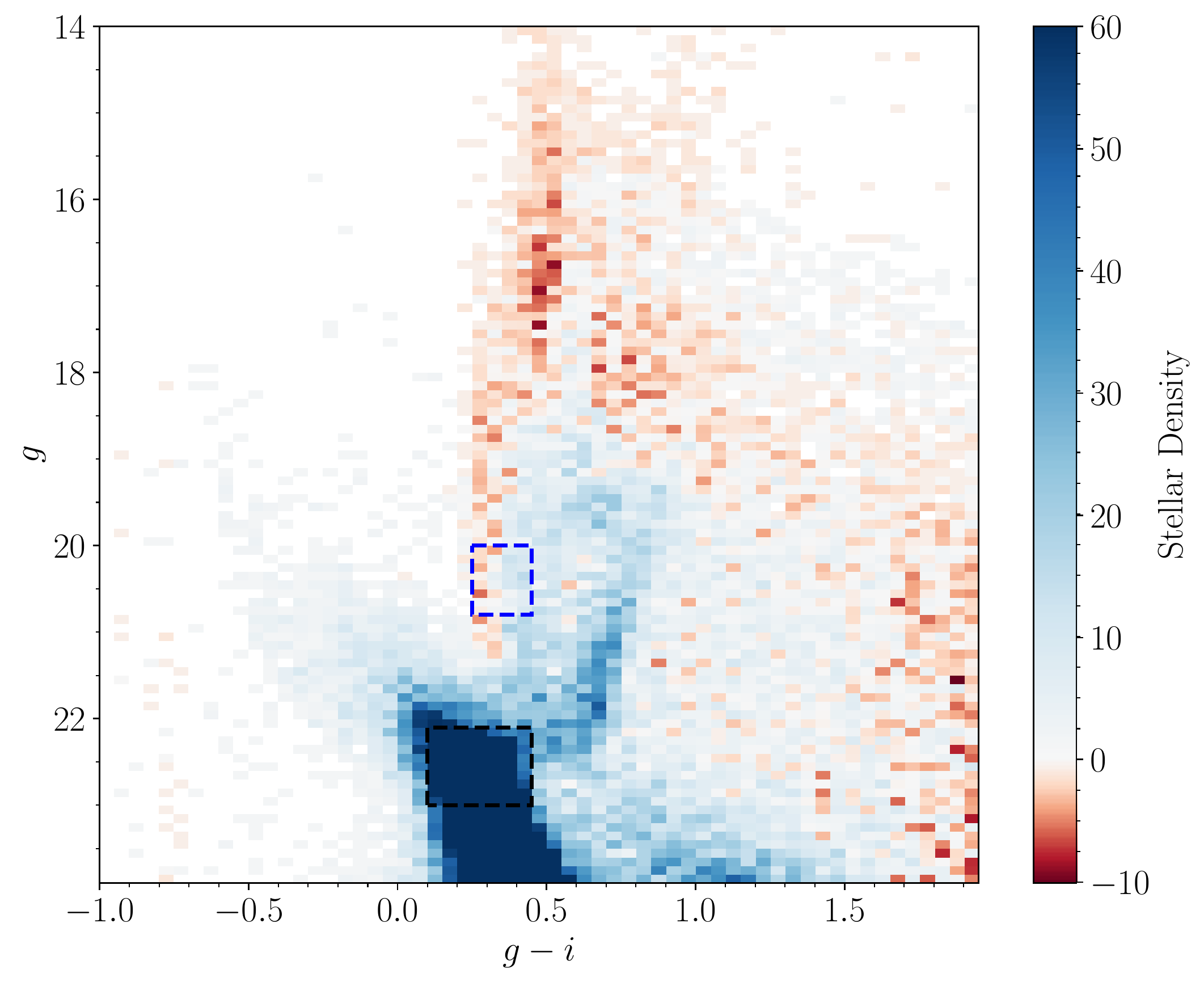}
    \caption{Hess diagram for field 13 after MW decontamination. The two dashed rectangles show the areas used to do the scaling of the halo of the MW to the SMASH data (blue) and the area around the oldest MSTO of the SMC, where the star count measurement is taken (black). The scaling area is set to avoid the SMC region and the MW dominated part ($g<19$ mag).}
    \label{fig:msto_cmd_modelzone}
\end{figure}

We used the same MW models described in Section \ref{subsec:MW_model}, but applying a different scaling factor than the one mentioned in Section \ref{subsec:MW_model}, given that the goal was to tailor our scaling to the outer parts of the system and to a very specific region of the CMD. The blue rectangle in Figure \ref{fig:msto_cmd_modelzone}, $0.25 \leq g-i \leq 0.45$ and $20 \leq g \leq 20.8$, represents an area of the CMD dominated by MW halo stars. By comparing the star counts in this area between the model and the SMASH data we are able to scale the model to each specific field. This method is not accurate in the inner parts of the SMC given that the scaling area in the CMD also contains SMC stars, creating a bias in the stellar count. Hence, we will focus our attention in those fields located at more than 4 degrees away from the centre of the SMC. 

The black dashed rectangle of Figure \ref{fig:msto_cmd_modelzone}, $0.1 \leq g-i \leq 0.45$ and $22.1 \leq g \leq 23$, shows the stars we adopted to calculate the density profile from the MSTO. This selection is less restrictive than that used by N19 because we want to include the effect of the larger distance spread along the line of sight of the SMC compared to the LMC. 

Instead of fitting a smooth profile for the SMC (as in Figure \ref{fig:SB_radial_profile}) here we aimed at  identifying individual fields that show some signs of possible SMC contributions at larger radii. For this  we needed an accurate estimate for any possible MW foreground stars or background galaxies remaining in the final sample used for calculation. This was done by taking 18 fields located further away than 12 degrees from the SMC centre that show no hints of contributions from either of the MCs, and then calculating the median value of stellar density and the 67\% confidence interval for each field.

\begin{figure*}
    \centering
    \includegraphics[width=\textwidth]{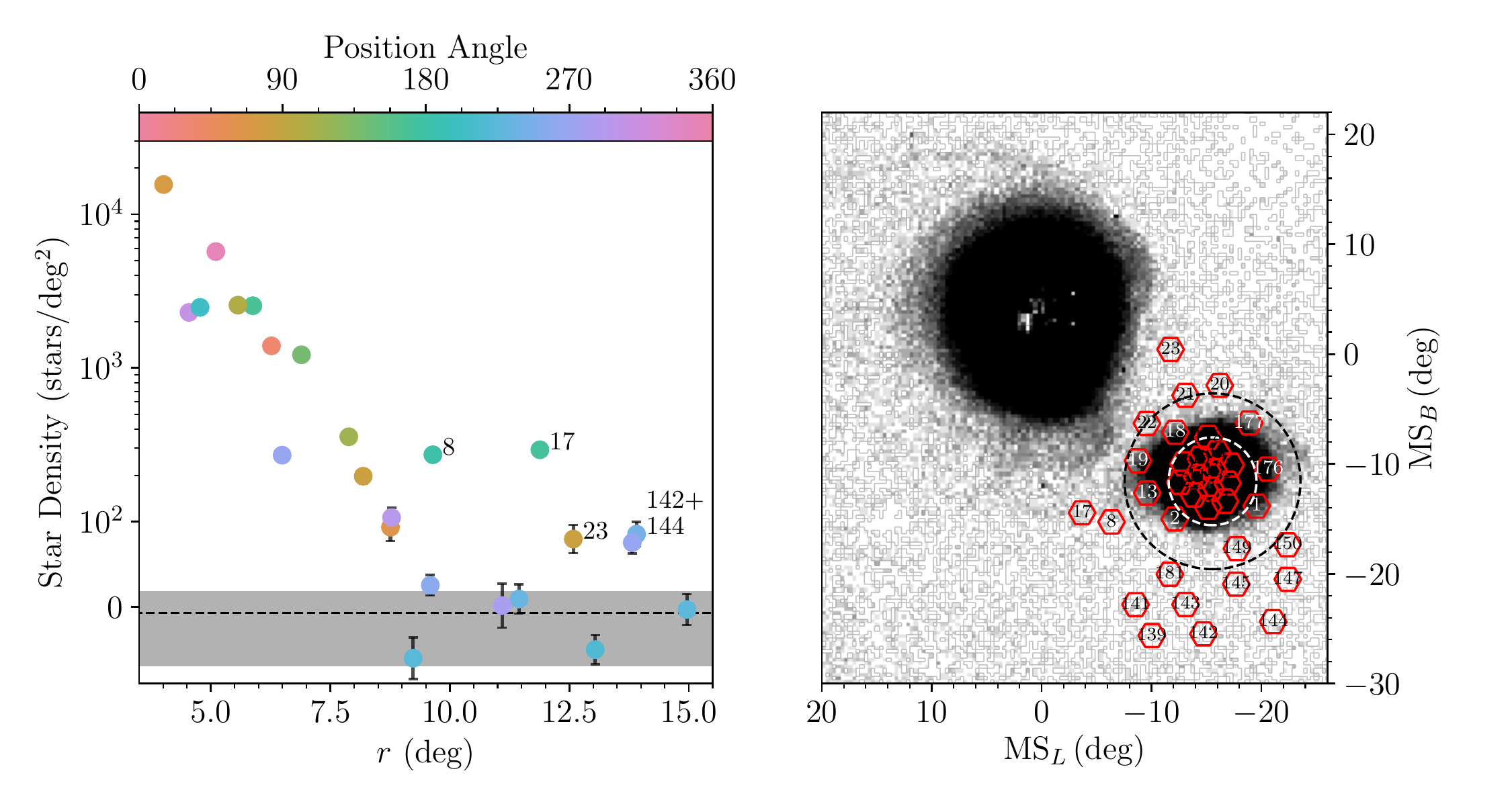}
    \caption{\textit{Left:} Star count density profile of the stars around the MSTO of the SMC. The profile starts at 3.5 degrees from the SMC centre with the aim of highlighting the detection at the very faint outskirts ($r > 7$ deg). The points are colour coded according to the PA of each field. The grey band shows our uncertainty in the foreground counts of the MW, and the error bars for each field represent the statistical Poisson uncertainty. We highlight the position of fields 8 and 17, that show a significantly higher density than the foreground level at radii larger than the expected for the SMC and as far as almost 12 degrees. \textit{Right:} Density map  of RGB stars, with saturation at 140 counts, from \citet{Belokurov2019}, with the SMASH fields overlayed as red polygons. The two dashed circles represent the 4 (white) and 8 (black) degree circles around the centre of the SMC. Fields 8 and 17 sit on an overdense region of Magellanic Cloud debris.}
    \label{fig:MSTO_profile_belokurov}
\end{figure*}

The new profile is shown in Figure \ref{fig:MSTO_profile_belokurov} (left) using the same PA convention as in Figure \ref{fig:SB_radial_profile}. We did not fit a profile to the data here; instead we show the calculated foreground MW contamination (grey band) with the method mentioned above. The error bars for each field represent the statistical Poisson uncertainty due to their number count that, in most cases, is too small to be seen. The position of fields 8 and 17 are highlighted in the profile since they present stellar densities of $\sim 250 \, \mathrm{stars} / \deg ^{2}$ above the shown foreground limit, at radii of $\sim 9.5$ and $\sim 12$ degrees, respectively. In the right panel of Figure \ref{fig:MSTO_profile_belokurov}, we compare the position of our fields with the sky map of \citet{Belokurov2019} using RGB stars in \textit{Gaia} DR2.  
Fields 8 and 17 fall on top of a Magellanic feature, between Magellanic Stream Longitudes of 0 and -10. This feature appears to be coming off the SMC in a stream-like fashion. We plan to further investigate the origin of this apparent tidal feature in a future analysis involving more accurate proper motions, combined with radial velocities and SFHs. 

\begin{figure}
    \centering
    \includegraphics[width=\columnwidth]{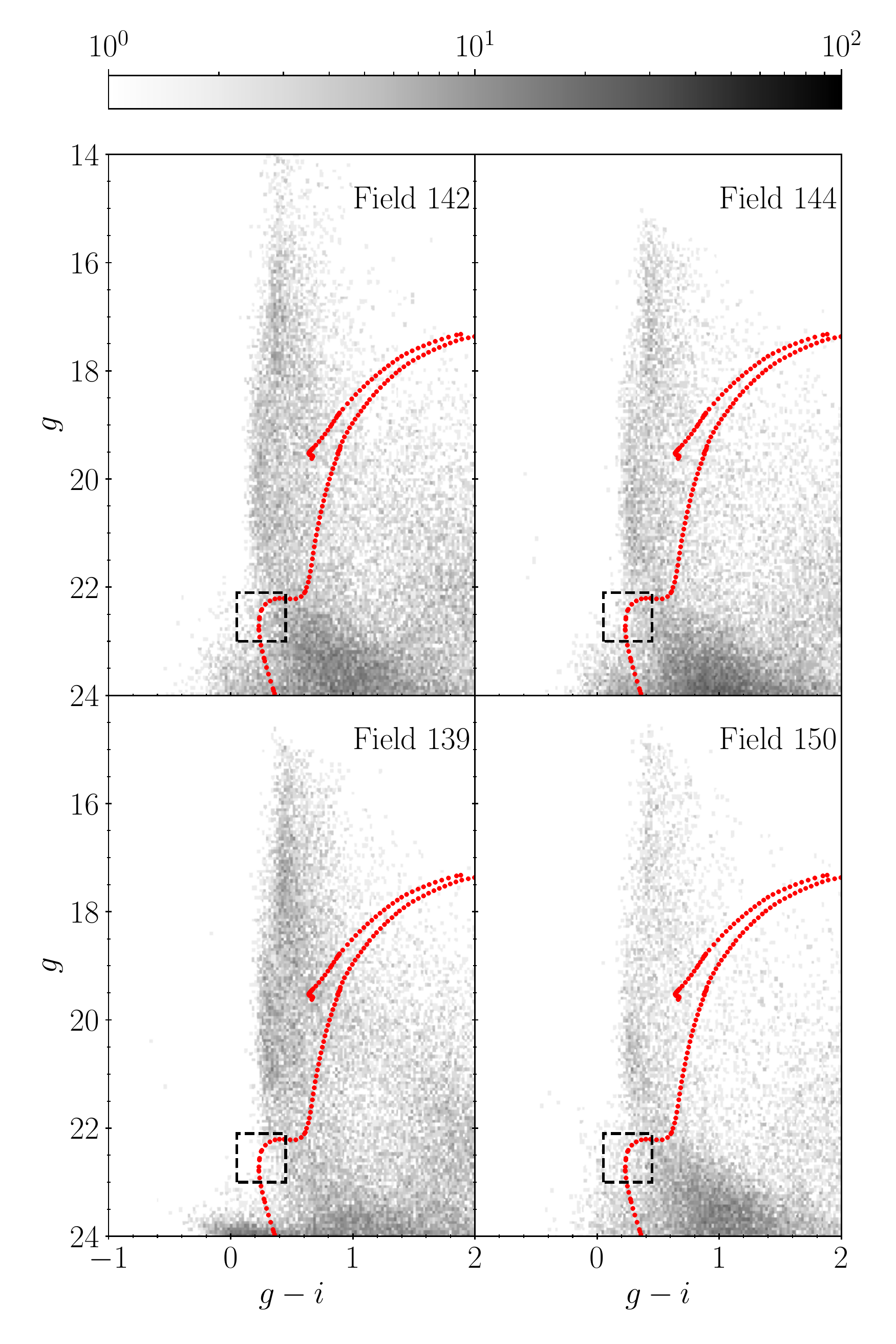}
    \caption{Hess diagrams of fields 142, 144, 139 and 150, without MW subtraction. The red isochrones represent a typical old SMC population (see Figure \ref{fig:example_cmds} for properties). The black box is signalling the area used to count stars around the MSTO (see profile of Figure \ref{fig:MSTO_profile_belokurov}). The top diagrams are meant to be compared with the bottom two. Field 139 shows no signs of either SMC or any further stellar population (other than MW halo) in the MSTO area. Field 150 shows some faint SMC stars. The other two may contain stars belonging to a structure further away from the MCs.}
    \label{fig:cmds_142_144}
\end{figure}

It is also worth mentioning field 23, located at $\sim 12.5 \degree$ and with a PA$\approx 90 \degree$, that has a stellar density that is slightly above the grey area in Figure \ref{fig:MSTO_profile_belokurov}. According to the RGB map, it is likely that this field contains LMC halo stars. Two additional fields present counts slightly above the grey line at $r \approx 14 \degree$:  fields 142 and 144.  Upon inspection of their CMDs (see Figure \ref{fig:cmds_142_144}), they do not appear to have any MCs debris.

We estimated the surface brightness of these features assuming that the stars in the MSTO area account for $\sim 10\%$ of the total luminosity of a stellar population with $\log \, (\mathrm{yr}) = 9.9$ and $[\mathrm{Fe/H}]=-1$, using PARSEC isochrones \citep{Bressan2012} with a \citet{Kroupa2002} initial mass function. The list of values obtained for all the different fields in Figure \ref{fig:MSTO_profile_belokurov} are also listed in Table \ref{tab:SB_profile}. Note that this method systematically underestimates the real value of the surface brightness ($\Sigma_g$) for fields with younger stellar populations due to the presence of more massive stars as is, for example, the case of fields 18 and 22. However, it is an effective method to infer the surface brightness of a typical old halo population.

For fields 8 and 17 we find similar surface brightness levels of $\Sigma_g = 31.91 \pm 0.09$ mag arcsec$^{-2}$ and $\Sigma_g = 31.92 \pm 0.07$ mag arcsec$^{-2}$, respectively. These are well above the limit found in N19 of $\Sigma_g \approx 34$ mag arcsec$^{-2}$ for low surface brightness features around the LMC, hinting that these are Magellanic debris and not something further away. Field 23 has a similar surface brightness to those of the faintest fields in N19, indicating that these are LMC debris. 

Finally, fields 142 and 144 constitute our faintest fields as computed with the MSTO density approach, with surface brightness of $\Sigma_g$ of 35.17 mag arcsec$^{-2}$ and 35.34 mag arcsec$^{-2}$, respectively. This is about one dex fainter than the faintest structure in N19 and, at the same time, have similar stellar density to fields 20, 23 and 150 that have Magellanic debris. This could be due to several factors. One of them could be the inclusion of unresolved galaxies contaminating the faint end of our MSTO selection box. This is unlikely because the selection of galaxies seems to work for the rest of the fields. The exposure times for these 2 fields are the same as well, indicating that the characteristics of the dataset in terms of uncertainties are the same. Another possibility is that these are part of a real structure at the locations of these fields. As can be seen in Figure \ref{fig:cmds_142_144}, the top panels correspond to the CMDs of these 2 fields and are compared to the bottom panels of a typical MW halo-only field (field 139) and one with traces of SMC material (field 150). Without knowing the nature of this putative feature it is difficult to make more conclusions and it falls beyond the scope of this paper.

\section{Disruption by the LMC} \label{sec:simulations}
To understand the stellar density distribution in the outskirts of the SMC better, it is useful to compare our results with simulations of the SMC's disruption. For this, we use numerical simulations similar to those described in \citet{Belokurov2017}. In particular, these simulations take the present 6D positions of the LMC and SMC, rewind them for 3 Gyr in the presence of each other and the MW, and then disrupt the SMC until the present day. The SMC's disruption is modelled using the Lagrange point stripping method of \citet{Gibbons2014}, that allows a rapid exploration of the parameter space given the observational uncertainties of the MCs present day positions and velocities. In these simulations, the LMC is modelled as a Hernquist profile, with a mass of $2.5\times10^{11}M_\odot$ and a scale radius of 25 kpc, and the SMC is modelled as a Plummer sphere, with a mass of $1\times10^{9} M_\odot$ and a scale radius of 1 kpc. Note that we are using a more massive SMC than that used in \citet{Belokurov2017} in order to explore the extended debris around the SMC, instead of the RR Lyrae overdensity  found in that work. This value has proved to match better the general spread of the debris.

\begin{figure}
\centering
\includegraphics[width=\columnwidth]{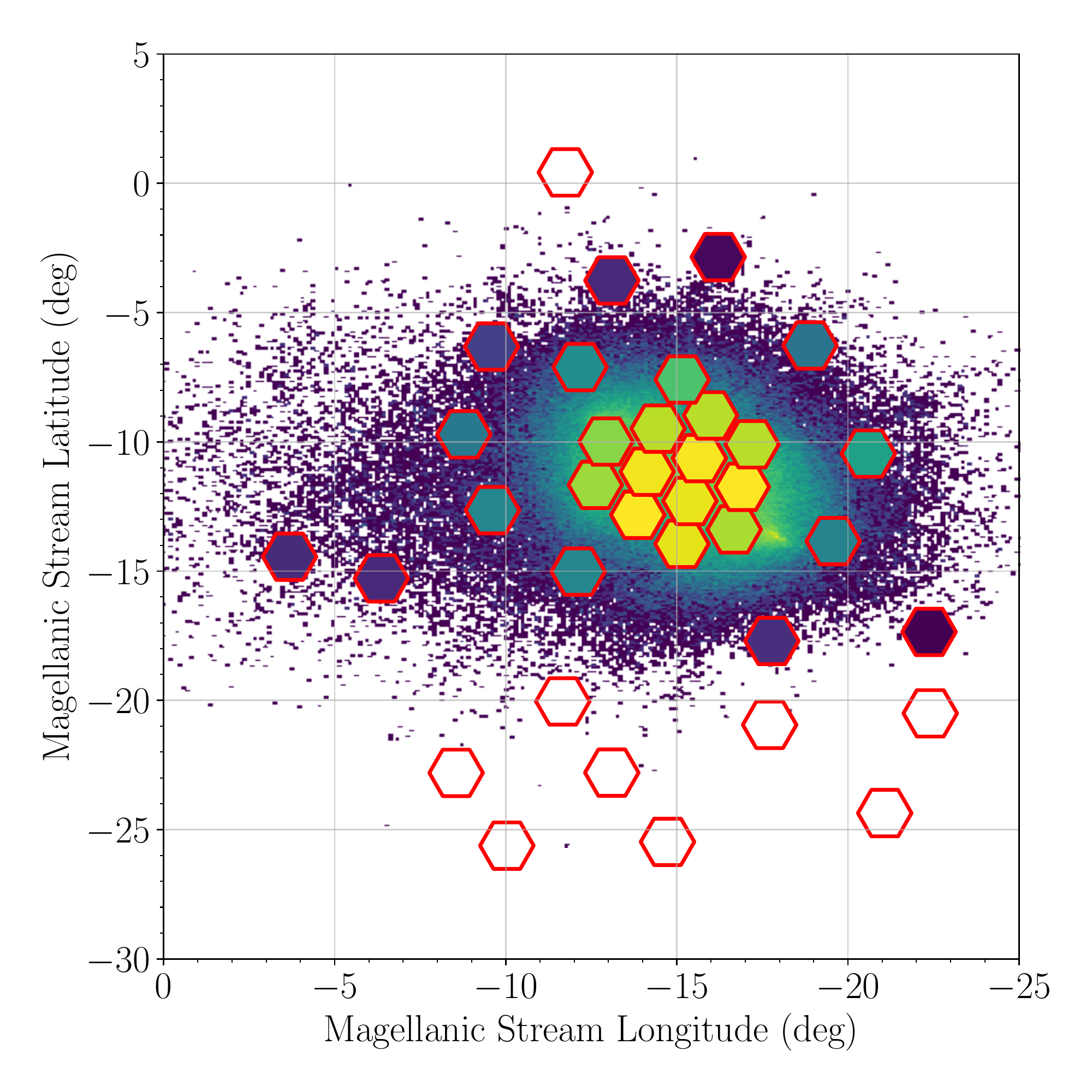}
\caption{Comparison between the tidal debris produced by our in-house simulation (background histogram) and the surface brightness extracted from the SMASH fields (foreground octagons). Only the fields with SMC stars have been coloured according to their MSTO stellar count. White octogons represent fields for which no SMC contribution has been detected according to the SB nor through a visual inspection of their CMDs.}
\label{fig:sim-vs-smash}
\end{figure}

Figure \ref{fig:sim-vs-smash} shows the distribution of the stellar debris of the SMC in the simulation compared to the number of total star counts that we have obtained in this work using SMASH. The octagons are colour-coded using the number of stellar counts in the SMASH fields. The CMDs of each individual field were analysed in detail and compared  to the surface brightness from Table \ref{tab:SB_profile} to determine the fields with the most obvious SMC contribution. Fields that are consistent with the MW foreground are then left white for visualisation purposes. The simulation is shown as a background 2-D histogram. To make sure that we are making realistic assumptions about the visibility of the debris around the galaxy, we only take stars stripped during the last episode of disruption in the galaxy around 250 Myr ago and represent them in Figure \ref{fig:sim-vs-smash}. 

The agreement between the extension of the debris and the SMASH fields with SMC stars is broadly consistent. This suggests that interactions with the LMC have helped to strip a substantial amount of material off the SMC, as previously shown by \citet{Carrera2017}. This, in turn, can also explain the elliptical shape of the outskirts of the SMC as the galaxy orbits around the LMC (\citealt{Nidever2011}, \citealt{Belokurov2017}, \citealt{Pieres2017}).

To test this, we calculated the tidal radius of the SMC due to the influence of the LMC. For that we used equation 7-84 from \citet{Binney2008}, but replacing the factor 3 with a 2 assuming the LMC has a flat rotation curve,

\begin{equation}
    r_t = d \left( \frac{M_{SMC}}{2 \, M_{LMC}} \right) ^{\frac{1}{3}}.
    \label{eq:tidal_radius}
\end{equation}

To calculate the distance between the SMC and LMC we use the latest distance measurements of $49.59$ kpc for the LMC \citep{Pietrzynski2019} and $63.4$ kpc for the SMC \citep{Ripepi2016}. For the masses, we use the estimate of the SMC from \cite{DiTeodoro2019} of $2.4\times10^9 $ M$_\odot$ (within $\sim$ 4 kpc) and use the value from \citet{Erkal2019b} of $1.38^{+0.27}_{-0.24} \times 10^{11} {M_\odot}$ for the LMC. With these values and the definition of on-sky centres by \citet{Paturel2003}, we obtained a separation between their centres of $\sim 24.3$ kpc, which then implies an SMC tidal radius of $r_t \approx 5.0$ kpc or $\approx 4.5$ degrees. This corresponds to a much smaller radius than that for which we find SMC stars, that could imply that the disruption of the SMC goes really far inside the profile of the galaxy as argued by \citet{DeLeo2020}. At the same time, all of the features described in the previous section would likely come from tidal disruption.

\section{Stellar populations structure} \label{sec: young vs. old}

A very interesting exercise is to look at the spatial distributions of stars of different ages. This was done in \citet{Zaritsky2000} for the internal parts of the SMC, where they show the highly irregular young population in contrast to the smooth older stellar population. More recent studies for both Clouds can be found in \citet{Youssoufi2019} and \citet{delPino2019}, using VMC and \textit{Gaia}, respectively. We know that interactions and subsequent disruption in galaxies can trigger episodes of star formation. The MCs are known to have ongoing star formation and a rich young stellar population across the main bodies and along the Magellanic Bridge. On the other hand, old stellar populations are good tracers of the formation and past evolution of the galaxy, because they are the primary constituents of the stellar halos of galaxies. By comparing the spatial distributions of young and old stellar populations we can extract information on the evolution of the galaxy, and, in particular, on the interactions with the LMC.

\begin{figure}
    \centering
    \includegraphics[width=\columnwidth]{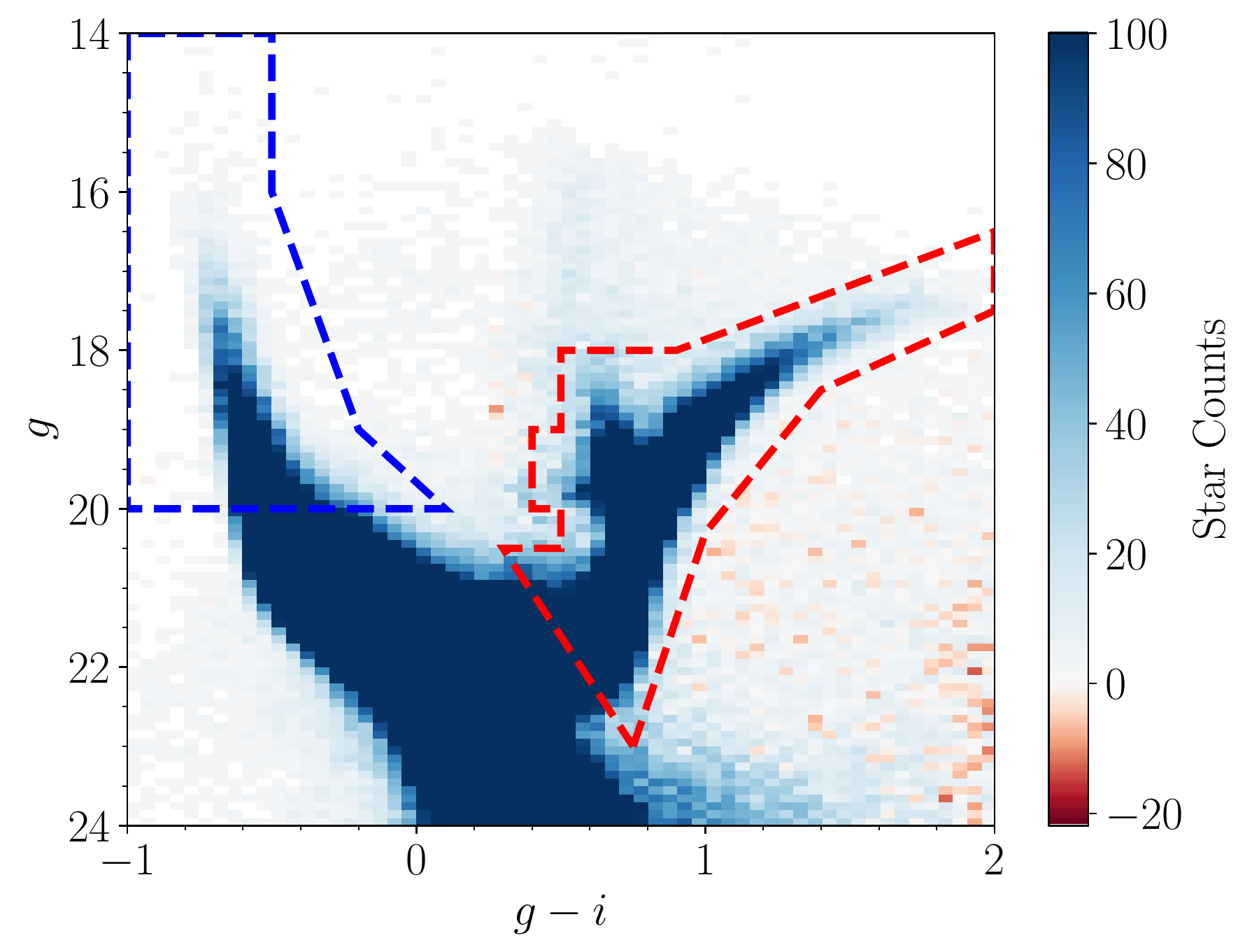}
    \caption{Hess diagram of SMASH field 3 showing two dashed contours that outline the regions used to calculate the on-sky densities of the two different stellar populations. The blue region contains younger ($ < 2 \, \mathrm{Gyr}$) SMC MS stars, while the red one encapsulates intermediate age to old stars, along with some contribution from the MW foreground.}
    \label{fig:pop_regions}
\end{figure}

For this purpose, we selected two different regions of the CMD that are known to be inhabited by stars of clearly different ages. This is shown in Figure \ref{fig:pop_regions} where we present the CMD for field 3 with two regions: the blue polygon representing the area where the young stars ($< 2$ Gyr) are taken from and the red polygon representing the area from which we extract the older population (making sure we include RGB and RC). Both regions are well separated from each other and avoid the lower MS part, where there is a degeneracy in the age of the stars. 

\begin{figure}
    \centering
    \includegraphics[width=\columnwidth]{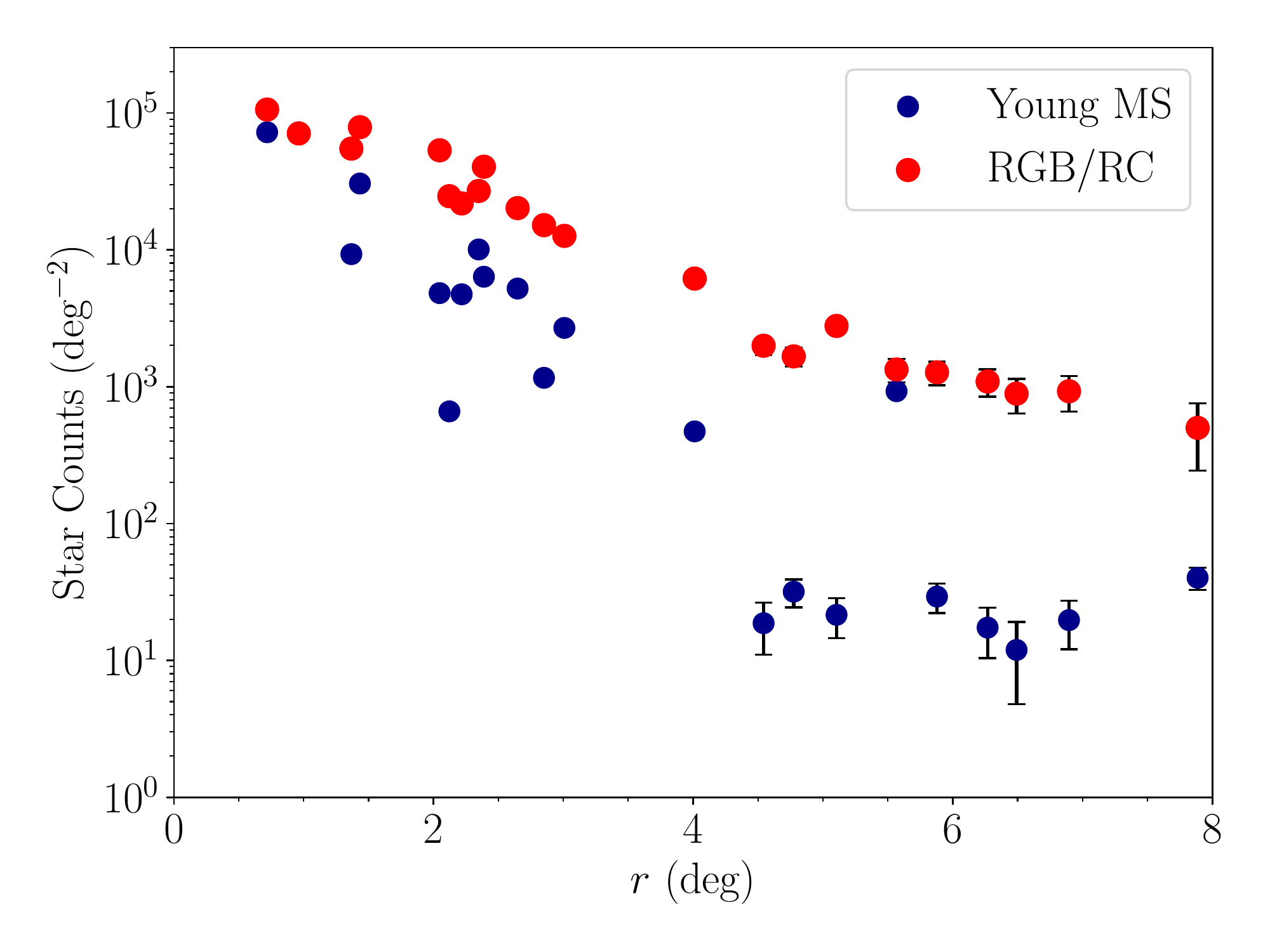}
    \caption{Star Count profile of the SMC as a function of distance for the two different populations inscribed by the regions shown in Figure \ref{fig:pop_regions}. The dark blue points represent the counts for the young main sequence population and the red points represent the counts for the RGB/RC population of (older) stars.}
    \label{fig:old_vs_young}
\end{figure}

Figure \ref{fig:old_vs_young} shows the star count profile for the two populations as a function of distance from the SMC centre. The younger population is depicted in blue and the older one in red. The bulk of the young stars is found in the inner 4 degrees of the galaxy, with a very irregular profile, as expected from the very perturbed gas reservoir of the galaxy \citep{Nidever2010}. After that, the profile flattens out with a very small contribution of these younger stars out to 8 degrees. The only exception to this is field 18 at a radius of $\sim 5.5$ degrees, which corresponds to a crowded region in the Magellanic Bridge with ongoing star formation. The older population presents a less disturbed profile and seems to dominate in the outer parts.

These distributions are consistent with the recent picture from the OGLE survey around the MCs \citep{Soszynski2019}, that also shows starkly different distributions between Cepheids (young) and RR Lyrae (old) stars. Similarly, \citet{Mackey2018} found prominent young stellar populations in the direction of the HI bridge between the Clouds. This is in agreement with our fields 18 (see Figure \ref{fig:example_cmds}) and 22 (last point at $\sim$8 degrees in Figure \ref{fig:old_vs_young}), which show very young star formation. \citet{Mackey2018} also show that the intermediate population ($\sim$1-4 Gyr) do not have the same distribution as the older population. Unfortunately this difference is not highlighted by our study.

To make this analysis more quantitative, we took the data for both populations and fitted them using the same approach as for the elliptical profile in Section \ref{sec:SB profile}. For the old/intermediate RGB population, we find a scale length of $h = 1.50 \pm 0.06$ degrees, much larger than that of the galaxy found previously in Section \ref{sec:SB profile}. We also find a lower S\'{e}rsic index of $n = 0.82 \pm 0.02$ and an almost negligible eccentricity $e = 0.05 \pm 0.01$. The young population follows a far more irregular profile, which is not fitted properly with either a circular or an elliptical shape. The values given by the MCMC algorithm show a steeper S\'{e}rsic profile with $n = 1.68 \pm 0.03$ and much more concentrated with $h = 0.107 \pm 0.008$ degrees, as well as showing a slightly higher eccentricity to that of the galaxy at $e = 0.183 \pm 0.005$. However, this fit does not reflect all the small scale structure of the young population and it is also not able to reproduce the young Magellanic Bridge feature. Even when this is not the appropriate treatment for the later population, combining these results with the view offered in Figure \ref{fig:old_vs_young}, we show that the young population is much more concentrated in the inner parts of the SMC.

\section{Conclusions} \label{sec:conclusions}

We conducted a careful analysis of 36 unprecedentedly deep CMDs of the SMC and its periphery obtained from images secured with the  DECam at the CTIO-4m Blanco telescope as part of the SMASH survey. We analysed the CMDs to determine the most accurate surface brightness profile of the SMC to date, constraining its extent with a careful decontamination process to remove MW foreground stars. \textit{Gaia} DR2 proper motion measurements were used to remove the foreground contamination by the  47 Tuc GC using the distinct proper motions between the SMC and the GC to disentangle the two populations. We  also measured the proper motions of both systems in the field of the cluster to be $(\mu_{\alpha} \cos{\delta}, \mu_{\delta}) = (0.547 \pm 0.002, -1.2453 \pm 0.0013)$ mas/yr for the SMC and $(\mu_{\alpha} \cos{\delta}, \mu_{\delta}) = (5.252 \pm 0.002, -2.494 \pm 0.002)$ mas/yr for 47 Tuc. To decontaminate the SMASH data from foreground NGC 362 stars, we used isochrone modelling, from which we reduced the uncertainty caused by this cluster in the total surface brightness of its respective field to 0.02\%. The rest of the decontamination process was done using mock catalogues of stars and it has been checked to yield consistent results throughout the fields.

The surface brightness profile of the SMC in polar coordinates shows a very staggered outline, i.e. high dispersion from the fit compared to the uncertainties, reaching as far out as 6-8 degrees depending on the direction.

We compared the 2-dimensional shape of this profile with simulations of a tidally stripped SMC with a mass of $\sim$ $10^9$ M$_\odot$ and found a similar debris extent. However, this mass represents the present-day mass,  indicating that the initial mass of the SMC is likely to have been larger than previously thought.  We interpret this as the result of the interactions with the LMC.

In addition to the elliptical shape, we identified a faint feature ($\sim 31.9$ mag arcsec$^{-2}$) present in SMASH fields 8 and 17, with a MSTO stellar density of $\sim 200\ \mathrm{stars}\, \deg ^{2}$ above the calculated limit of the foreground population. This feature has recently also been traced using RGB stars in \textit{Gaia} DR2 data \citep{Belokurov2019}. Its morphology is puzzling and a more contiguous coverage of the region, at a similar depth, is required to evaluate its shape properly.

Additionally, we find two fields (142 and 144) with high values of stellar density that we hypothesise might be showing the presence of some kind of structure  in the MW halo, despite not being able to make robust conclusions about its nature.

Finally, we explored the relative distributions of young and intermediate/old stars. 
The older populations show evidence of continuous and smooth profile, with an almost circular structure with scale-length $h = 1.50 \pm 0.06$ and shallow S\'{e}rsic index.
However, the young population presents a jump in the density at around 4 degrees from the SMC centre, and then a flat profile up to 8 degrees from the centre. 
We find that the young stellar part of the SMC has a scale-length of $h = 0.107 \pm 0.008$. This length is shorter than that of the total stellar surface brightness profile. This broken profile suggests that the SMC contents are heavily disturbed, showing that the processes affecting the outer HI morphology leave an imprint on the young stellar component of the SMC.

\section*{Acknowledgements}

We thank the anonymous referee for their insightful comments which helped us in expressing more clarity of our analysis and made the paper more robust. P.M. and N.E.D.N. would like to thank the hospitality of the Astronomy and Astrophysics Group at the Montana State University. S.R.M. acknowledges funding from grant AST-1909497 from the National Science Foundation. A.M.  acknowledges support from FONDECYT Regular grant 1181797 and funding from the Max Planck Society through a Partner Group grant. N.F.M. gratefully acknowledge support from the French National Research Agency (ANR) funded project ``Pristine'' (ANR-18-CE31-0017) along with funding from CNRS/INSU through the Programme National Galaxies et Cosmologie. M.-R.C. and C.P.M.B. acknowledge support from the European Research Council (ERC) under the European Union's Horizon 2020 research and innovation programme (grant agreement no. 682115). R.R.M. acknowledges partial support from project BASAL AFB-$170002$ as well as FONDECYT project N$^{\circ}1170364$. C.G., T.R.L., and M.M. acknowledge financial support through the grants (AEI/FEDER, UE) AYA2017-89076-P, and AYA2015-63810-P, as well as by the Ministerio de Ciencia, Innovaci\'on y Universidades (MCIU), through the State Budget and by the Consejer\'\i a de Econom\'\i a, Industria, Comercio y Conocimiento of the Canary Islands Autonomous Community, through the Regional Budget (including IAC project, TRACES). T.R.L. is also supported by grant AYA2016-77237-C3-1-P (RAVET project) and a MCIU Juan de la Cierva - Formaci\'on grant (FJCI-2016-30342). D.M.D. acknowledges financial support from the State Agency for Research of the Spanish MCIU through the ``Centre of Excellence Severo Ochoa" award for the Instituto de Astrof\'{i}sica de Andaluc\'{i}a (SEV-2017-0709).

Based on observations at Cerro Tololo Inter-American Observatory, NSF's National Optical-Infrared Astronomy Research Laboratory (NOIRLab Prop. ID: 2013A-0411 and 2013B-0440; PI: Nidever), which is operated by the Association of Universities for Research in Astronomy (AURA) under a cooperative agreement with the National Science Foundation. This project used data obtained with the Dark Energy Camera (DECam), which was constructed by the Dark Energy Survey (DES) collaborating institutions: Argonne National Lab, University of California Santa Cruz, University of Cambridge, Centro de Investigaciones Energ\'eticas, Medioambientales y Tecnol\'ogicas-Madrid, University of Chicago, University College London, DES-Brazil consortium, University of Edinburgh, ETH-Zurich, University of Illinois at Urbana-Champaign, Institut de Ci\`{e}ncies de l'Espai, Institut de F\'isica d'Altes Energies, Lawrence Berkeley National Lab, Ludwig-Maximilians Universit\"{a}t, University of Michigan, National Optical Astronomy Observatory, University of Nottingham, Ohio State University, University of Pennsylvania, University of Portsmouth, SLAC National Lab, Stanford University, University of Sussex, and Texas A\&M University. Funding for DES, including DECam, has been provided by the U.S. Department of Energy, National Science Foundation, Ministry of Education and Science (Spain), Science and Technology Facilities Council (UK), Higher Education Funding Council (England), National Center for Supercomputing Applications, Kavli Institute for Cosmological Physics, Financiadora de Estudos e Projetos, Funda\c{c}\~{a}o Carlos Chagas Filho de Amparo a Pesquisa, Conselho Nacional de Desenvolvimento Cient\'ifico e Tecnol\'ogico and the Minist\'erio da Ci\^encia e Tecnologia (Brazil), the German Research Foundation-sponsored cluster of excellence ``Origin and Structure of the Universe" and the DES collaborating institutions.

This research made use of Astropy,\footnote{http://www.astropy.org} a community-developed core Python package for Astronomy \citep{astropy:2013, astropy:2018}.



\bibliographystyle{mnras}
\bibliography{references}

\begin{thebibliography}{}
\makeatletter
\relax
\def\mn@urlcharsother{\let\do\@makeother \do\$\do\&\do\#\do\^\do\_\do\%\do\~}
\def\mn@doi{\begingroup\mn@urlcharsother \@ifnextchar [ {\mn@doi@}
  {\mn@doi@[]}}
\def\mn@doi@[#1]#2{\def\@tempa{#1}\ifx\@tempa\@empty \href
  {http://dx.doi.org/#2} {doi:#2}\else \href {http://dx.doi.org/#2} {#1}\fi
  \endgroup}
\def\mn@eprint#1#2{\mn@eprint@#1:#2::\@nil}
\def\mn@eprint@arXiv#1{\href {http://arxiv.org/abs/#1} {{\tt arXiv:#1}}}
\def\mn@eprint@dblp#1{\href {http://dblp.uni-trier.de/rec/bibtex/#1.xml}
  {dblp:#1}}
\def\mn@eprint@#1:#2:#3:#4\@nil{\def\@tempa {#1}\def\@tempb {#2}\def\@tempc
  {#3}\ifx \@tempc \@empty \let \@tempc \@tempb \let \@tempb \@tempa \fi \ifx
  \@tempb \@empty \def\@tempb {arXiv}\fi \@ifundefined
  {mn@eprint@\@tempb}{\@tempb:\@tempc}{\expandafter \expandafter \csname
  mn@eprint@\@tempb\endcsname \expandafter{\@tempc}}}

\bibitem[\protect\citeauthoryear{{Abbott} et~al.,}{{Abbott}
  et~al.}{2018}]{Abbott2018}
{Abbott} T.~M.~C.,  et~al., 2018, \mn@doi [\apjs] {10.3847/1538-4365/aae9f0},
  \href {https://ui.adsabs.harvard.edu/abs/2018ApJS..239...18A} {239, 18}

\bibitem[\protect\citeauthoryear{{Astropy Collaboration} et~al.,}{{Astropy
  Collaboration} et~al.}{2013}]{astropy:2013}
{Astropy Collaboration} et~al., 2013, \mn@doi [\aap]
  {10.1051/0004-6361/201322068}, \href
  {http://adsabs.harvard.edu/abs/2013A%26A...558A..33A} {558, A33}

\bibitem[\protect\citeauthoryear{{Bekki} \& {Chiba}}{{Bekki} \&
  {Chiba}}{2005}]{Bekki2005}
{Bekki} K.,  {Chiba} M.,  2005, \mn@doi [\mnras]
  {10.1111/j.1365-2966.2004.08510.x}, \href
  {https://ui.adsabs.harvard.edu/#abs/2005MNRAS.356..680B} {356, 680}

\bibitem[\protect\citeauthoryear{{Bell} et~al.,}{{Bell}
  et~al.}{2008}]{Bell2008}
{Bell} E.~F.,  et~al., 2008, \mn@doi [\apj] {10.1086/588032}, \href
  {https://ui.adsabs.harvard.edu/abs/2008ApJ...680..295B} {680, 295}

\bibitem[\protect\citeauthoryear{{Bellini}, {Bianchini}, {Varri}, {Anderson},
  {Piotto}, {van der Marel}, {Vesperini}  \& {Watkins}}{{Bellini}
  et~al.}{2017}]{Bellini2017}
{Bellini} A.,  {Bianchini} P.,  {Varri} A.~L.,  {Anderson} J.,  {Piotto} G.,
  {van der Marel} R.~P.,  {Vesperini} E.,   {Watkins} L.~L.,  2017, \mn@doi
  [\apj] {10.3847/1538-4357/aa7c5f}, \href
  {https://ui.adsabs.harvard.edu/abs/2017ApJ...844..167B} {844, 167}

\bibitem[\protect\citeauthoryear{{Belokurov} \& {Erkal}}{{Belokurov} \&
  {Erkal}}{2019}]{Belokurov2019}
{Belokurov} V.~A.,  {Erkal} D.,  2019, \mn@doi [\mnras]
  {10.1093/mnrasl/sly178}, \href
  {https://ui.adsabs.harvard.edu/abs/2019MNRAS.482L...9B} {482, L9}

\bibitem[\protect\citeauthoryear{{Belokurov}, {Erkal}, {Deason}, {Koposov}, {De
  Angeli}, {Evans}, {Fraternali}  \& {Mackey}}{{Belokurov}
  et~al.}{2017}]{Belokurov2017}
{Belokurov} V.,  {Erkal} D.,  {Deason} A.~J.,  {Koposov} S.~E.,  {De Angeli}
  F.,  {Evans} D.~W.,  {Fraternali} F.,   {Mackey} D.,  2017, \mn@doi [\mnras]
  {10.1093/mnras/stw3357}, \href
  {https://ui.adsabs.harvard.edu/#abs/2017MNRAS.466.4711B} {466, 4711}

\bibitem[\protect\citeauthoryear{{Besla}, {Kallivayalil}, {Hernquist},
  {Robertson}, {Cox}, {van der Marel}  \& {Alcock}}{{Besla}
  et~al.}{2007}]{Besla2007B}
{Besla} G.,  {Kallivayalil} N.,  {Hernquist} L.,  {Robertson} B.,  {Cox} T.~J.,
   {van der Marel} R.~P.,   {Alcock} C.,  2007, \mn@doi [\apj]
  {10.1086/521385}, \href
  {https://ui.adsabs.harvard.edu/#abs/2007ApJ...668..949B} {668, 949}

\bibitem[\protect\citeauthoryear{{Besla}, {Kallivayalil}, {Hernquist}, {van der
  Marel}, {Cox}  \& {Kere{\v{s}}}}{{Besla} et~al.}{2012}]{Besla2012}
{Besla} G.,  {Kallivayalil} N.,  {Hernquist} L.,  {van der Marel} R.~P.,  {Cox}
  T.~J.,   {Kere{\v{s}}} D.,  2012, \mn@doi [\mnras]
  {10.1111/j.1365-2966.2012.20466.x}, \href
  {https://ui.adsabs.harvard.edu/#abs/2012MNRAS.421.2109B} {421, 2109}

\bibitem[\protect\citeauthoryear{{Besla}, {Mart{\'\i}nez-Delgado}, {van der
  Marel}, {Beletsky}, {Seibert}, {Schlafly}, {Grebel}  \& {Neyer}}{{Besla}
  et~al.}{2016}]{Besla2016}
{Besla} G.,  {Mart{\'\i}nez-Delgado} D.,  {van der Marel} R.~P.,  {Beletsky}
  Y.,  {Seibert} M.,  {Schlafly} E.~F.,  {Grebel} E.~K.,   {Neyer} F.,  2016,
  \mn@doi [\apj] {10.3847/0004-637X/825/1/20}, \href
  {https://ui.adsabs.harvard.edu/#abs/2016ApJ...825...20B} {825, 20}

\bibitem[\protect\citeauthoryear{{Binney} \& {Tremaine}}{{Binney} \&
  {Tremaine}}{2008}]{Binney2008}
{Binney} J.,  {Tremaine} S.,  2008, {Galactic Dynamics: Second Edition}

\bibitem[\protect\citeauthoryear{{Bressan}, {Marigo}, {Girardi}, {Salasnich},
  {Dal Cero}, {Rubele}  \& {Nanni}}{{Bressan} et~al.}{2012}]{Bressan2012}
{Bressan} A.,  {Marigo} P.,  {Girardi} L.,  {Salasnich} B.,  {Dal Cero} C.,
  {Rubele} S.,   {Nanni} A.,  2012, \mn@doi [\mnras]
  {10.1111/j.1365-2966.2012.21948.x}, \href
  {https://ui.adsabs.harvard.edu/abs/2012MNRAS.427..127B} {427, 127}

\bibitem[\protect\citeauthoryear{{Carrera}, {Gallart}, {Hardy}, {Aparicio}  \&
  {Zinn}}{{Carrera} et~al.}{2008a}]{Carrera2008a}
{Carrera} R.,  {Gallart} C.,  {Hardy} E.,  {Aparicio} A.,   {Zinn} R.,  2008a,
  \mn@doi [\aj] {10.1088/0004-6256/135/3/836}, \href
  {https://ui.adsabs.harvard.edu/abs/2008AJ....135..836C} {135, 836}

\bibitem[\protect\citeauthoryear{{Carrera}, {Gallart}, {Aparicio}, {Costa},
  {M{\'e}ndez}  \& {No{\"e}l}}{{Carrera} et~al.}{2008b}]{Carrera2008b}
{Carrera} R.,  {Gallart} C.,  {Aparicio} A.,  {Costa} E.,  {M{\'e}ndez} R.~A.,
   {No{\"e}l} N. E.~D.,  2008b, \mn@doi [\aj] {10.1088/0004-6256/136/3/1039},
  \href {https://ui.adsabs.harvard.edu/#abs/2008AJ....136.1039C} {136, 1039}

\bibitem[\protect\citeauthoryear{{Carrera}, {Conn}, {No{\"e}l}, {Read}  \&
  {L{\'o}pez S{\'a}nchez}}{{Carrera} et~al.}{2017}]{Carrera2017}
{Carrera} R.,  {Conn} B.~C.,  {No{\"e}l} N. E.~D.,  {Read} J.~I.,   {L{\'o}pez
  S{\'a}nchez} {\'A}.~R.,  2017, \mn@doi [\mnras] {10.1093/mnras/stx1932},
  \href {https://ui.adsabs.harvard.edu/#abs/2017MNRAS.471.4571C} {471, 4571}

\bibitem[\protect\citeauthoryear{{Chen}, {Richer}, {Caiazzo}  \& {Heyl}}{{Chen}
  et~al.}{2018}]{Chen2018}
{Chen} S.,  {Richer} H.,  {Caiazzo} I.,   {Heyl} J.,  2018, \mn@doi [\apj]
  {10.3847/1538-4357/aae089}, \href
  {https://ui.adsabs.harvard.edu/\#abs/2018ApJ...867..132C} {867, 132}

\bibitem[\protect\citeauthoryear{{Choi}, {Dotter}, {Conroy}, {Cantiello},
  {Paxton}  \& {Johnson}}{{Choi} et~al.}{2016}]{Choi2016}
{Choi} J.,  {Dotter} A.,  {Conroy} C.,  {Cantiello} M.,  {Paxton} B.,
  {Johnson} B.~D.,  2016, \mn@doi [\apj] {10.3847/0004-637X/823/2/102}, \href
  {https://ui.adsabs.harvard.edu/abs/2016ApJ...823..102C} {823, 102}

\bibitem[\protect\citeauthoryear{{Choi} et~al.,}{{Choi}
  et~al.}{2018a}]{Choi2018a}
{Choi} Y.,  et~al., 2018a, \mn@doi [\apj] {10.3847/1538-4357/aae083}, \href
  {https://ui.adsabs.harvard.edu/abs/2018ApJ...866...90C} {866, 90}

\bibitem[\protect\citeauthoryear{{Choi} et~al.,}{{Choi}
  et~al.}{2018b}]{Choi2018b}
{Choi} Y.,  et~al., 2018b, \mn@doi [\apj] {10.3847/1538-4357/aaed1f}, \href
  {https://ui.adsabs.harvard.edu/abs/2018ApJ...869..125C} {869, 125}

\bibitem[\protect\citeauthoryear{{Cioni} et~al.,}{{Cioni}
  et~al.}{2011}]{Cioni2011}
{Cioni} M. R.~L.,  et~al., 2011, \mn@doi [\aap] {10.1051/0004-6361/201016137},
  \href {https://ui.adsabs.harvard.edu/abs/2011A&A...527A.116C} {527, A116}

\bibitem[\protect\citeauthoryear{{De Leo}, {Carrera}, {Noel}, {Read}, {Erkal}
  \& {Gallart}}{{De Leo} et~al.}{2020}]{DeLeo2020}
{De Leo} M.,  {Carrera} R.,  {Noel} N. E.~D.,  {Read} J.~I.,  {Erkal} D.,
  {Gallart} C.,  2020, arXiv e-prints, \href
  {https://ui.adsabs.harvard.edu/abs/2020arXiv200211138D} {p. arXiv:2002.11138}

\bibitem[\protect\citeauthoryear{{Deason}, {Belokurov}  \& {Evans}}{{Deason}
  et~al.}{2011}]{Deason2011}
{Deason} A.~J.,  {Belokurov} V.,   {Evans} N.~W.,  2011, \mn@doi [\mnras]
  {10.1111/j.1365-2966.2011.19237.x}, \href
  {https://ui.adsabs.harvard.edu/abs/2011MNRAS.416.2903D} {416, 2903}

\bibitem[\protect\citeauthoryear{{Di Teodoro} et~al.,}{{Di Teodoro}
  et~al.}{2019}]{DiTeodoro2019}
{Di Teodoro} E.~M.,  et~al., 2019, \mn@doi [\mnras] {10.1093/mnras/sty3095},
  \href {https://ui.adsabs.harvard.edu/abs/2019MNRAS.483..392D} {483, 392}

\bibitem[\protect\citeauthoryear{{El Youssoufi} et~al.,}{{El Youssoufi}
  et~al.}{2019}]{Youssoufi2019}
{El Youssoufi} D.,  et~al., 2019, \mn@doi [\mnras] {10.1093/mnras/stz2400},
  \href {https://ui.adsabs.harvard.edu/abs/2019MNRAS.490.1076E} {490, 1076}

\bibitem[\protect\citeauthoryear{{Erkal} et~al.,}{{Erkal}
  et~al.}{2019}]{Erkal2019b}
{Erkal} D.,  et~al., 2019, \mn@doi [\mnras] {10.1093/mnras/stz1371}, \href
  {https://ui.adsabs.harvard.edu/abs/2019MNRAS.tmp.1318E} {p.~1318}

\bibitem[\protect\citeauthoryear{{Flaugher}}{{Flaugher}}{2005}]{Flaugher2005}
{Flaugher} B.,  2005, \mn@doi [International Journal of Modern Physics A]
  {10.1142/S0217751X05025917}, \href
  {https://ui.adsabs.harvard.edu/abs/2005IJMPA..20.3121F} {20, 3121}

\bibitem[\protect\citeauthoryear{{Flaugher} et~al.,}{{Flaugher}
  et~al.}{2015}]{Flaugher2015}
{Flaugher} B.,  et~al., 2015, \mn@doi [\aj] {10.1088/0004-6256/150/5/150},
  \href {https://ui.adsabs.harvard.edu/#abs/2015AJ....150..150F} {150, 150}

\bibitem[\protect\citeauthoryear{{Foreman-Mackey}, {Hogg}, {Lang}  \&
  {Goodman}}{{Foreman-Mackey} et~al.}{2013}]{Foreman-Mackey2013}
{Foreman-Mackey} D.,  {Hogg} D.~W.,  {Lang} D.,   {Goodman} J.,  2013, \mn@doi
  [Publications of the Astronomical Society of the Pacific] {10.1086/670067},
  \href {https://ui.adsabs.harvard.edu/\#abs/2013PASP..125..306F} {125, 306}

\bibitem[\protect\citeauthoryear{{Gaia Collaboration} et~al.,}{{Gaia
  Collaboration} et~al.}{2018}]{Gaia2018A}
{Gaia Collaboration} et~al., 2018, \mn@doi [\aap]
  {10.1051/0004-6361/201833051}, \href
  {https://ui.adsabs.harvard.edu/abs/2018A&A...616A...1G} {616, A1}

\bibitem[\protect\citeauthoryear{{Gallart}, {Stetson}, {Hardy}, {Pont}  \&
  {Zinn}}{{Gallart} et~al.}{2004}]{Gallart2004}
{Gallart} C.,  {Stetson} P.~B.,  {Hardy} E.,  {Pont} F.,   {Zinn} R.,  2004,
  \mn@doi [\apj] {10.1086/425866}, \href
  {https://ui.adsabs.harvard.edu/#abs/2004ApJ...614L.109G} {614, L109}

\bibitem[\protect\citeauthoryear{{Gallart} et~al.,}{{Gallart}
  et~al.}{2015}]{Gallart2015}
{Gallart} C.,  et~al., 2015, \mn@doi [\apjl] {10.1088/2041-8205/811/2/L18},
  \href {https://ui.adsabs.harvard.edu/abs/2015ApJ...811L..18G} {811, L18}

\bibitem[\protect\citeauthoryear{{Gardiner} \& {Hawkins}}{{Gardiner} \&
  {Hawkins}}{1991}]{Gardiner1991}
{Gardiner} L.~T.,  {Hawkins} M.~R.~S.,  1991, \mn@doi [\mnras]
  {10.1093/mnras/251.1.174}, \href
  {https://ui.adsabs.harvard.edu/#abs/1991MNRAS.251..174G} {251, 174}

\bibitem[\protect\citeauthoryear{{Gibbons}, {Belokurov}  \& {Evans}}{{Gibbons}
  et~al.}{2014}]{Gibbons2014}
{Gibbons} S.~L.~J.,  {Belokurov} V.,   {Evans} N.~W.,  2014, \mn@doi [\mnras]
  {10.1093/mnras/stu1986}, \href
  {https://ui.adsabs.harvard.edu/#abs/2014MNRAS.445.3788G} {445, 3788}

\bibitem[\protect\citeauthoryear{{Glatt} et~al.,}{{Glatt}
  et~al.}{2008}]{Glatt2008}
{Glatt} K.,  et~al., 2008, \mn@doi [\aj] {10.1088/0004-6256/135/4/1106}, \href
  {https://ui.adsabs.harvard.edu/abs/2008AJ....135.1106G} {135, 1106}

\bibitem[\protect\citeauthoryear{{Grogin} et~al.,}{{Grogin}
  et~al.}{2011}]{Grogin2011}
{Grogin} N.~A.,  et~al., 2011, \mn@doi [\apjs] {10.1088/0067-0049/197/2/35},
  \href {https://ui.adsabs.harvard.edu/abs/2011ApJS..197...35G} {197, 35}

\bibitem[\protect\citeauthoryear{{Harmsen}, {Monachesi}, {Bell}, {de Jong},
  {Bailin}, {Radburn-Smith}  \& {Holwerda}}{{Harmsen}
  et~al.}{2017}]{Harmsen2017}
{Harmsen} B.,  {Monachesi} A.,  {Bell} E.~F.,  {de Jong} R.~S.,  {Bailin} J.,
  {Radburn-Smith} D.~J.,   {Holwerda} B.~W.,  2017, \mn@doi [\mnras]
  {10.1093/mnras/stw2992}, \href
  {https://ui.adsabs.harvard.edu/abs/2017MNRAS.466.1491H} {466, 1491}

\bibitem[\protect\citeauthoryear{{Hindman}, {Kerr}  \& {McGee}}{{Hindman}
  et~al.}{1963}]{Hindman1963}
{Hindman} J.~V.,  {Kerr} F.~J.,   {McGee} R.~X.,  1963, \mn@doi [Australian
  Journal of Physics] {10.1071/PH630570}, \href
  {https://ui.adsabs.harvard.edu/#abs/1963AuJPh..16..570H} {16, 570}

\bibitem[\protect\citeauthoryear{{Kallivayalil}, {van der Marel}, {Alcock},
  {Axelrod}, {Cook}, {Drake}  \& {Geha}}{{Kallivayalil}
  et~al.}{2006}]{Kallivayalil2006a}
{Kallivayalil} N.,  {van der Marel} R.~P.,  {Alcock} C.,  {Axelrod} T.,  {Cook}
  K.~H.,  {Drake} A.~J.,   {Geha} M.,  2006, \mn@doi [\apj] {10.1086/498972},
  \href {https://ui.adsabs.harvard.edu/#abs/2006ApJ...638..772K} {638, 772}

\bibitem[\protect\citeauthoryear{{Kallivayalil}, {van der Marel}, {Besla},
  {Anderson}  \& {Alcock}}{{Kallivayalil} et~al.}{2013}]{Kallivayalil2013}
{Kallivayalil} N.,  {van der Marel} R.~P.,  {Besla} G.,  {Anderson} J.,
  {Alcock} C.,  2013, \mn@doi [\apj] {10.1088/0004-637X/764/2/161}, \href
  {https://ui.adsabs.harvard.edu/#abs/2013ApJ...764..161K} {764, 161}

\bibitem[\protect\citeauthoryear{{Kallivayalil} et~al.,}{{Kallivayalil}
  et~al.}{2018}]{Kallivayalil2018}
{Kallivayalil} N.,  et~al., 2018, \mn@doi [\apj] {10.3847/1538-4357/aadfee},
  \href {https://ui.adsabs.harvard.edu/abs/2018ApJ...867...19K} {867, 19}

\bibitem[\protect\citeauthoryear{{Kroupa}}{{Kroupa}}{2002}]{Kroupa2002}
{Kroupa} P.,  2002, \mn@doi [Science] {10.1126/science.1067524}, \href
  {https://ui.adsabs.harvard.edu/abs/2002Sci...295...82K} {295, 82}

\bibitem[\protect\citeauthoryear{{Mackey}, {Koposov}, {Da Costa}, {Belokurov},
  {Erkal}  \& {Kuzma}}{{Mackey} et~al.}{2018}]{Mackey2018}
{Mackey} D.,  {Koposov} S.,  {Da Costa} G.,  {Belokurov} V.,  {Erkal} D.,
  {Kuzma} P.,  2018, \mn@doi [\apj] {10.3847/2041-8213/aac175}, \href
  {https://ui.adsabs.harvard.edu/#abs/2018ApJ...858L..21M} {858, L21}

\bibitem[\protect\citeauthoryear{{Martin} et~al.,}{{Martin}
  et~al.}{2015}]{Martin2015}
{Martin} N.~F.,  et~al., 2015, \mn@doi [\apjl] {10.1088/2041-8205/804/1/L5},
  \href {https://ui.adsabs.harvard.edu/abs/2015ApJ...804L...5M} {804, L5}

\bibitem[\protect\citeauthoryear{{Martin} et~al.,}{{Martin}
  et~al.}{2016}]{Martin2016a}
{Martin} N.~F.,  et~al., 2016, \mn@doi [\apjl] {10.3847/2041-8205/830/1/L10},
  \href {http://adsabs.harvard.edu/abs/2016ApJ...830L..10M} {830, L10}

\bibitem[\protect\citeauthoryear{{Mart{\'\i}nez-Delgado}
  et~al.,}{{Mart{\'\i}nez-Delgado} et~al.}{2019}]{Martinez-Delgado2019}
{Mart{\'\i}nez-Delgado} D.,  et~al., 2019, \mn@doi [\aap]
  {10.1051/0004-6361/201936021}, \href
  {https://ui.adsabs.harvard.edu/abs/2019A&A...631A..98M} {631, A98}

\bibitem[\protect\citeauthoryear{{Mathewson}, {Cleary}  \&
  {Murray}}{{Mathewson} et~al.}{1974}]{Mathewson1974}
{Mathewson} D.~S.,  {Cleary} M.~N.,   {Murray} J.~D.,  1974, \mn@doi [\apj]
  {10.1086/152875}, \href
  {https://ui.adsabs.harvard.edu/#abs/1974ApJ...190..291M} {190, 291}

\bibitem[\protect\citeauthoryear{{Mau} et~al.,}{{Mau} et~al.}{2020}]{Mau2020}
{Mau} S.,  et~al., 2020, \mn@doi [\apj] {10.3847/1538-4357/ab6c67}, \href
  {https://ui.adsabs.harvard.edu/abs/2020ApJ...890..136M} {890, 136}

\bibitem[\protect\citeauthoryear{{McConnachie} et~al.,}{{McConnachie}
  et~al.}{2009}]{McConnachie2009}
{McConnachie} A.~W.,  et~al., 2009, \mn@doi [\nat] {10.1038/nature08327}, \href
  {https://ui.adsabs.harvard.edu/#abs/2009Natur.461...66M} {461, 66}

\bibitem[\protect\citeauthoryear{{Mu{\~n}oz} et~al.,}{{Mu{\~n}oz}
  et~al.}{2006}]{Munoz2006a}
{Mu{\~n}oz} R.~R.,  et~al., 2006, \mn@doi [\apj] {10.1086/505620}, \href
  {https://ui.adsabs.harvard.edu/#abs/2006ApJ...649..201M} {649, 201}

\bibitem[\protect\citeauthoryear{{Muraveva} et~al.,}{{Muraveva}
  et~al.}{2018}]{Muraveva2018}
{Muraveva} T.,  et~al., 2018, \mn@doi [\mnras] {10.1093/mnras/stx2514}, \href
  {https://ui.adsabs.harvard.edu/#abs/2018MNRAS.473.3131M} {473, 3131}

\bibitem[\protect\citeauthoryear{{Nidever}, {Majewski}, {Butler Burton}  \&
  {Nigra}}{{Nidever} et~al.}{2010}]{Nidever2010}
{Nidever} D.~L.,  {Majewski} S.~R.,  {Butler Burton} W.,   {Nigra} L.,  2010,
  \mn@doi [\apj] {10.1088/0004-637X/723/2/1618}, \href
  {https://ui.adsabs.harvard.edu/#abs/2010ApJ...723.1618N} {723, 1618}

\bibitem[\protect\citeauthoryear{{Nidever}, {Majewski}, {Mu{\~n}oz}, {Beaton},
  {Patterson}  \& {Kunkel}}{{Nidever} et~al.}{2011}]{Nidever2011}
{Nidever} D.~L.,  {Majewski} S.~R.,  {Mu{\~n}oz} R.~R.,  {Beaton} R.~L.,
  {Patterson} R.~J.,   {Kunkel} W.~E.,  2011, \mn@doi [\apj]
  {10.1088/2041-8205/733/1/L10}, \href
  {https://ui.adsabs.harvard.edu/#abs/2011ApJ...733L..10N} {733, L10}

\bibitem[\protect\citeauthoryear{{Nidever}, {Monachesi}, {Bell}, {Majewski},
  {Mu{\~n}oz}  \& {Beaton}}{{Nidever} et~al.}{2013}]{Nidever2013}
{Nidever} D.~L.,  {Monachesi} A.,  {Bell} E.~F.,  {Majewski} S.~R.,
  {Mu{\~n}oz} R.~R.,   {Beaton} R.~L.,  2013, \mn@doi [\apj]
  {10.1088/0004-637X/779/2/145}, \href
  {http://adsabs.harvard.edu/abs/2013ApJ...779..145N} {779, 145}

\bibitem[\protect\citeauthoryear{{Nidever} et~al.,}{{Nidever}
  et~al.}{2017}]{Nidever2017}
{Nidever} D.~L.,  et~al., 2017, \mn@doi [\aj] {10.3847/1538-3881/aa8d1c}, \href
  {http://adsabs.harvard.edu/abs/2017AJ....154..199N} {154, 199}

\bibitem[\protect\citeauthoryear{{Nidever} et~al.,}{{Nidever}
  et~al.}{2019}]{Nidever2019}
{Nidever} D.~L.,  et~al., 2019, \mn@doi [\apj] {10.3847/1538-4357/aafaf7},
  \href {https://ui.adsabs.harvard.edu/abs/2019ApJ...874..118N} {874, 118}

\bibitem[\protect\citeauthoryear{{Niederhofer} et~al.,}{{Niederhofer}
  et~al.}{2018}]{Niederhofer2018a}
{Niederhofer} F.,  et~al., 2018, \mn@doi [\aap] {10.1051/0004-6361/201732144},
  \href {https://ui.adsabs.harvard.edu/abs/2018A&A...612A.115N} {612, A115}

\bibitem[\protect\citeauthoryear{{No{\"e}l} \& {Gallart}}{{No{\"e}l} \&
  {Gallart}}{2007}]{Noel2007B}
{No{\"e}l} N.~E.~D.,  {Gallart} C.,  2007, \mn@doi [\apjl] {10.1086/521223},
  \href {http://adsabs.harvard.edu/abs/2007ApJ...665L..23N} {665, L23}

\bibitem[\protect\citeauthoryear{{No{\"e}l}, {Gallart}, {Costa}  \&
  {M{\'e}ndez}}{{No{\"e}l} et~al.}{2007}]{Noel2007A}
{No{\"e}l} N. E.~D.,  {Gallart} C.,  {Costa} E.,   {M{\'e}ndez} R.~A.,  2007,
  \mn@doi [\aj] {10.1086/512668}, \href
  {https://ui.adsabs.harvard.edu/abs/2007AJ....133.2037N} {133, 2037}

\bibitem[\protect\citeauthoryear{{No{\"e}l}, {Aparicio}, {Gallart}, {Hidalgo},
  {Costa}  \& {M{\'e}ndez}}{{No{\"e}l} et~al.}{2009}]{Noel2009}
{No{\"e}l} N. E.~D.,  {Aparicio} A.,  {Gallart} C.,  {Hidalgo} S.~L.,  {Costa}
  E.,   {M{\'e}ndez} R.~A.,  2009, \mn@doi [\apj]
  {10.1088/0004-637X/705/2/1260}, \href
  {https://ui.adsabs.harvard.edu/abs/2009ApJ...705.1260N} {705, 1260}

\bibitem[\protect\citeauthoryear{{No{\"e}l}, {Conn}, {Carrera}, {Read}, {Rix}
  \& {Dolphin}}{{No{\"e}l} et~al.}{2013}]{Noel2013a}
{No{\"e}l} N.~E.~D.,  {Conn} B.~C.,  {Carrera} R.,  {Read} J.~I.,  {Rix} H.~W.,
    {Dolphin} A.,  2013, \mn@doi [\apj] {10.1088/0004-637X/768/2/109}, \href
  {https://ui.adsabs.harvard.edu/#abs/2013ApJ...768..109N} {768, 109}

\bibitem[\protect\citeauthoryear{{No{\"e}l}, {Conn}, {Read}, {Carrera},
  {Dolphin}  \& {Rix}}{{No{\"e}l} et~al.}{2015}]{Noel2015}
{No{\"e}l} N.~E.~D.,  {Conn} B.~C.,  {Read} J.~I.,  {Carrera} R.,  {Dolphin}
  A.,   {Rix} H.~W.,  2015, \mn@doi [\mnras] {10.1093/mnras/stv1614}, \href
  {https://ui.adsabs.harvard.edu/#abs/2015MNRAS.452.4222N} {452, 4222}

\bibitem[\protect\citeauthoryear{{Olsen}, {Zaritsky}, {Blum}, {Boyer}  \&
  {Gordon}}{{Olsen} et~al.}{2011}]{Olsen2011}
{Olsen} K. A.~G.,  {Zaritsky} D.,  {Blum} R.~D.,  {Boyer} M.~L.,   {Gordon}
  K.~D.,  2011, \mn@doi [\apj] {10.1088/0004-637X/737/1/29}, \href
  {https://ui.adsabs.harvard.edu/abs/2011ApJ...737...29O} {737, 29}

\bibitem[\protect\citeauthoryear{{Paturel}, {Petit}, {Prugniel}, {Theureau},
  {Rousseau}, {Brouty}, {Dubois}  \& {Cambr{\'e}sy}}{{Paturel}
  et~al.}{2003}]{Paturel2003}
{Paturel} G.,  {Petit} C.,  {Prugniel} P.,  {Theureau} G.,  {Rousseau} J.,
  {Brouty} M.,  {Dubois} P.,   {Cambr{\'e}sy} L.,  2003, \mn@doi [\aap]
  {10.1051/0004-6361:20031411}, \href
  {https://ui.adsabs.harvard.edu/abs/2003A&A...412...45P} {412, 45}

\bibitem[\protect\citeauthoryear{{Pe{\~n}arrubia}, {G{\'o}mez}, {Besla},
  {Erkal}  \& {Ma}}{{Pe{\~n}arrubia} et~al.}{2016}]{Penarrubia:2016}
{Pe{\~n}arrubia} J.,  {G{\'o}mez} F.~A.,  {Besla} G.,  {Erkal} D.,   {Ma}
  Y.-Z.,  2016, \mn@doi [\mnras] {10.1093/mnrasl/slv160}, \href
  {https://ui.adsabs.harvard.edu/abs/2016MNRAS.456L..54P} {456, L54}

\bibitem[\protect\citeauthoryear{{Piatti}, {de Grijs}, {Rubele}, {Cioni},
  {Ripepi}  \& {Kerber}}{{Piatti} et~al.}{2015}]{Piatti2015}
{Piatti} A.~E.,  {de Grijs} R.,  {Rubele} S.,  {Cioni} M.-R.~L.,  {Ripepi} V.,
   {Kerber} L.,  2015, \mn@doi [\mnras] {10.1093/mnras/stv635}, \href
  {https://ui.adsabs.harvard.edu/#abs/2015MNRAS.450..552P} {450, 552}

\bibitem[\protect\citeauthoryear{{Pieres} et~al.,}{{Pieres}
  et~al.}{2017}]{Pieres2017}
{Pieres} A.,  et~al., 2017, \mn@doi [\mnras] {10.1093/mnras/stx507}, \href
  {https://ui.adsabs.harvard.edu/#abs/2017MNRAS.468.1349P} {468, 1349}

\bibitem[\protect\citeauthoryear{{Pietrzy{\'n}ski} et~al.,}{{Pietrzy{\'n}ski}
  et~al.}{2019}]{Pietrzynski2019}
{Pietrzy{\'n}ski} G.,  et~al., 2019, \mn@doi [\nat]
  {10.1038/s41586-019-0999-4}, \href
  {https://ui.adsabs.harvard.edu/abs/2019Natur.567..200P} {567, 200}

\bibitem[\protect\citeauthoryear{{Price-Whelan} et~al.,}{{Price-Whelan}
  et~al.}{2018}]{astropy:2018}
{Price-Whelan} A.~M.,  et~al., 2018, \mn@doi [\aj] {10.3847/1538-3881/aabc4f},
  \href {https://ui.adsabs.harvard.edu/#abs/2018AJ....156..123T} {156, 123}

\bibitem[\protect\citeauthoryear{{Putman} et~al.,}{{Putman}
  et~al.}{1998}]{Putman1998}
{Putman} M.~E.,  et~al., 1998, \mn@doi [\nat] {10.1038/29466}, \href
  {https://ui.adsabs.harvard.edu/#abs/1998Natur.394..752P} {394, 752}

\bibitem[\protect\citeauthoryear{{Read}, {Pontzen}  \& {Viel}}{{Read}
  et~al.}{2006}]{Read2006c}
{Read} J.~I.,  {Pontzen} A.~P.,   {Viel} M.,  2006, \mn@doi [\mnras]
  {10.1111/j.1365-2966.2006.10720.x}, \href
  {https://ui.adsabs.harvard.edu/#abs/2006MNRAS.371..885R} {371, 885}

\bibitem[\protect\citeauthoryear{{Ripepi} et~al.,}{{Ripepi}
  et~al.}{2016}]{Ripepi2016}
{Ripepi} V.,  et~al., 2016, \mn@doi [The Astrophysical Journal Supplement
  Series] {10.3847/0067-0049/224/2/21}, \href
  {https://ui.adsabs.harvard.edu/#abs/2016ApJS..224...21R} {224, 21}

\bibitem[\protect\citeauthoryear{{Ripepi} et~al.,}{{Ripepi}
  et~al.}{2017}]{Ripepi2017}
{Ripepi} V.,  et~al., 2017, \mn@doi [\mnras] {10.1093/mnras/stx2096}, \href
  {https://ui.adsabs.harvard.edu/#abs/2017MNRAS.472..808R} {472, 808}

\bibitem[\protect\citeauthoryear{{Robin}, {Reyl{\'e}}, {Derri{\`e}re}  \&
  {Picaud}}{{Robin} et~al.}{2003}]{Robin2003}
{Robin} A.~C.,  {Reyl{\'e}} C.,  {Derri{\`e}re} S.,   {Picaud} S.,  2003,
  \mn@doi [\aap] {10.1051/0004-6361:20031117}, \href
  {https://ui.adsabs.harvard.edu/abs/2003A&A...409..523R} {409, 523}

\bibitem[\protect\citeauthoryear{{Schlafly} \& {Finkbeiner}}{{Schlafly} \&
  {Finkbeiner}}{2011}]{Schlafly2011}
{Schlafly} E.~F.,  {Finkbeiner} D.~P.,  2011, \mn@doi [\apj]
  {10.1088/0004-637X/737/2/103}, \href
  {https://ui.adsabs.harvard.edu/abs/2011ApJ...737..103S} {737, 103}

\bibitem[\protect\citeauthoryear{{Schlegel}, {Finkbeiner}  \&
  {Davis}}{{Schlegel} et~al.}{1998}]{Schlegel1998}
{Schlegel} D.~J.,  {Finkbeiner} D.~P.,   {Davis} M.,  1998, \mn@doi [\apj]
  {10.1086/305772}, \href
  {https://ui.adsabs.harvard.edu/abs/1998ApJ...500..525S} {500, 525}

\bibitem[\protect\citeauthoryear{{Scowcroft}, {Freedman}, {Madore}, {Monson},
  {Persson}, {Rich}, {Seibert}  \& {Rigby}}{{Scowcroft}
  et~al.}{2016}]{Scowcroft2016}
{Scowcroft} V.,  {Freedman} W.~L.,  {Madore} B.~F.,  {Monson} A.,  {Persson}
  S.~E.,  {Rich} J.,  {Seibert} M.,   {Rigby} J.~R.,  2016, \mn@doi [\apj]
  {10.3847/0004-637X/816/2/49}, \href
  {https://ui.adsabs.harvard.edu/#abs/2016ApJ...816...49S} {816, 49}

\bibitem[\protect\citeauthoryear{{Sharma}, {Bland-Hawthorn}, {Johnston}  \&
  {Binney}}{{Sharma} et~al.}{2011}]{Sharma2011}
{Sharma} S.,  {Bland-Hawthorn} J.,  {Johnston} K.~V.,   {Binney} J.,  2011,
  \mn@doi [\apj] {10.1088/0004-637X/730/1/3}, \href
  {https://ui.adsabs.harvard.edu/abs/2011ApJ...730....3S} {730, 3}

\bibitem[\protect\citeauthoryear{{Skrutskie} et~al.,}{{Skrutskie}
  et~al.}{2006}]{Skrutskie2006}
{Skrutskie} M.~F.,  et~al., 2006, \mn@doi [\aj] {10.1086/498708}, \href
  {https://ui.adsabs.harvard.edu/#abs/2006AJ....131.1163S} {131, 1163}

\bibitem[\protect\citeauthoryear{{Slater}, {Nidever}, {Munn}, {Bell}  \&
  {Majewski}}{{Slater} et~al.}{2016}]{Slater2016}
{Slater} C.~T.,  {Nidever} D.~L.,  {Munn} J.~A.,  {Bell} E.~F.,   {Majewski}
  S.~R.,  2016, \mn@doi [\apj] {10.3847/0004-637X/832/2/206}, \href
  {https://ui.adsabs.harvard.edu/abs/2016ApJ...832..206S} {832, 206}

\bibitem[\protect\citeauthoryear{{Soszy{\'n}ski} et~al.,}{{Soszy{\'n}ski}
  et~al.}{2019}]{Soszynski2019}
{Soszy{\'n}ski} I.,  et~al., 2019, \mn@doi [\actaa]
  {10.32023/0001-5237/69.2.1}, \href
  {https://ui.adsabs.harvard.edu/abs/2019AcA....69...87S} {69, 87}

\bibitem[\protect\citeauthoryear{{Subramanian} et~al.,}{{Subramanian}
  et~al.}{2017}]{Subramanian2017}
{Subramanian} S.,  et~al., 2017, \mn@doi [\mnras] {10.1093/mnras/stx205}, \href
  {https://ui.adsabs.harvard.edu/#abs/2017MNRAS.467.2980S} {467, 2980}

\bibitem[\protect\citeauthoryear{{Zaritsky}, {Harris}, {Grebel}  \&
  {Thompson}}{{Zaritsky} et~al.}{2000}]{Zaritsky2000}
{Zaritsky} D.,  {Harris} J.,  {Grebel} E.~K.,   {Thompson} I.~B.,  2000,
  \mn@doi [\apjl] {10.1086/312649}, \href
  {https://ui.adsabs.harvard.edu/abs/2000ApJ...534L..53Z} {534, L53}

\bibitem[\protect\citeauthoryear{{Zivick} et~al.,}{{Zivick}
  et~al.}{2018}]{Zivick2018}
{Zivick} P.,  et~al., 2018, \mn@doi [\apj] {10.3847/1538-4357/aad4b0}, \href
  {https://ui.adsabs.harvard.edu/abs/2018ApJ...864...55Z} {864, 55}

\bibitem[\protect\citeauthoryear{{de Grijs} \& {Bono}}{{de Grijs} \&
  {Bono}}{2015}]{DeGrijs2015}
{de Grijs} R.,  {Bono} G.,  2015, \mn@doi [\aj] {10.1088/0004-6256/149/6/179},
  \href {https://ui.adsabs.harvard.edu/#abs/2015AJ....149..179D} {149, 179}

\bibitem[\protect\citeauthoryear{{del Pino}}{{del Pino}}{2019}]{delPino2019}
{del Pino} A.,  2019, in A Synoptic View of the Magellanic Clouds: VMC, Gaia
  and Beyond. p.~39, \mn@doi{10.5281/zenodo.3472522}

\makeatother
\end{thebibliography}

\vspace{0.3in}
\noindent\textit{
$^{1}$Department of Physics, University of Surrey, Guildford, GU2 7XH, UK \\
$^{2}$Department of Physics, Montana State University, P.O. Box 173840, Bozeman, MT 59717-3840, USA \\
$^{3}$Institute for Astronomy, University of Hawai'i, 2680 Woodlawn Drive, Honolulu, HI 96822, USA \\
$^{4}$Space Telescope Science Institute, 3700 San Martin Drive, Baltimore, MD 21218, USA \\
$^{5}$Department of Astronomy, University of Virginia, Charlottesville, VA 22904, USA \\
$^{6}$NSF's National Optical-Infrared Astronomy Research Laboratory, 950 N. Cherry Ave., Tucson, AZ 85719, USA \\
$^{7}$Instituto de Investigaci\'{o}n Multidisciplinario en Ciencia y Tecnolog\'{i}a, Universidad de La Serena, Ra\'{u}l Bitr\'{a}n 1305, La Serena, Chile \\
$^{8}$Departamento de F\'{i}sica y Astronom\'{i}a, Universidad de La Serena, Av. Juan Cisternas 1200 N, La Serena, Chile \\
$^{9}$Instituto de Astrof\'{i}sica de Canarias, Calle V\'{i}a L\'{a}ctea s/n, E-38205 La Laguna, Tenerife, Spain \\
$^{10}$Departamento de Astrof\'{i}sica, Universidad de La Laguna, E-38200 La Laguna, Tenerife, Spain \\
$^{11}$Center for Astrophysical Sciences, Department of Physics \& Astronomy, Johns Hopkins University, Baltimore, MD 21218, USA \\
$^{12}$Steward Observatory, University of Arizona, 933 North Cherry Avenue, Tucson, AZ 85721, USA \\
$^{13}$Observatoire astronomique de Strasbourg, Universit\'{e} de Strasbourg, CNRS, UMR 7550, 11 rue de l'Universit\'{e}, F-67000 Strasbourg, France \\
$^{14}$Max-Planck-Institut f\"{u}r Astronomie, K\"{o}nigstuhl 17, D-69117, Heidelberg, Germany \\
$^{15}$Departamento de Astronom\'{i}a, Universidad de Chile, Camino del Observatorio 1515, Las Condes, Santiago, Chile \\
$^{16}$Leibniz-Institut f\"{u}r Astrophysik Potsdam (AIP), An der Sternwarte 16, D-14482 Potsdam, Germany \\
$^{17}$Department of Astronomy, University of Michigan, 1085 S. University Avenue, Ann Arbor, MI 48109-1107, USA \\
$^{18}$Center for Astrophysics and Space Astronomy, University of Colorado, 389 UCB, Boulder, CO 80309-0389, USA \\
$^{19}$Institute of Astronomy, Madingley Rd, Cambridge, CB3 0HA \\
$^{20}$Cerro Tololo Inter-American Observatory, NSF's National Optical-Infrared Astronomy Research Laboratory, Casilla 603, La Serena, Chile \\
$^{21}$Instituto de Astrof\'{i}sica de Andaluc\'{i}a, Glorieta de la Astronom\'{i}a s/n, 18008, Granada, Spain \\
$^{22}$Insurance Australia Group Limited, Level 13, Tower Two Darling Park 201 Sussex Street Sydney NSW 2000
}







\bsp	
\label{lastpage}
\end{document}